\def\rey{$Re$}

\def\etal{et al.}

\documentclass[%
 aip,
 amsmath,amssymb,
preprint,%
floatfix,
]{revtex4-1}

\usepackage{graphicx}
\usepackage{dcolumn}
\usepackage{bm}
\usepackage[utf8]{inputenc}
\usepackage[T1]{fontenc}
\usepackage{mathptmx}
\usepackage{xcolor}
\usepackage[normalem]{ulem}
\usepackage{hyperref}

\begin{document}

\title[]{Aerodynamic interaction of bristled wing pairs in fling}

\author{Vishwa T.~Kasoju}
\author{Arvind Santhanakrishnan}%
 \email{askrish@okstate.edu}
\affiliation{School of Mechanical and Aerospace Engineering, Oklahoma State University,\\Stillwater, OK 74078, USA}

\date{\today}

\begin{abstract}
Tiny flying insects of body lengths under 2 mm use the `clap-and-fling' mechanism with bristled wings for lift augmentation and drag reduction at chord-based Reynolds number (\rey) on $\mathcal{O}(10)$. We examine wing-wing interaction of bristled wings in fling at \rey=10, as a function of initial inter-wing spacing ($\delta$) and degree of overlap between rotation and linear translation. A dynamically scaled robotic platform was used to drive physical models of bristled wing pairs with the following kinematics (all angles relative to vertical): 1) rotation about the trailing edge to angle $\theta_\text{r}$; 2) linear translation at a fixed angle ($\theta_\text{t}$); and 3) combined rotation and linear translation. The results show that: 1) cycle-averaged drag coefficient decreased with increasing $\theta_\text{r}$ and $\theta_\text{t}$; and 2) decreasing $\delta$ increased the lift coefficient owing to increased asymmetry in circulation of leading and trailing edge vortices. A new dimensionless index, reverse flow capacity (RFC), was used to quantify the maximum possible ability of a bristled wing to leak fluid through the bristles. Drag coefficients were larger for smaller $\delta$ and $\theta_\text{r}$ despite larger RFC, likely due to blockage of inter-bristle flow by shear layers around the bristles. Smaller $\delta$ during early rotation resulted in formation of strong positive pressure distribution between the wings, resulting in increased drag force. The positive pressure region weakened with increasing $\theta_\text{r}$, which in turn reduced drag force. Tiny insects have been previously reported to use large rotational angles in fling, and our findings suggest that a plausible reason is to reduce drag forces.\\

\noindent{\bf\em The following article has been submitted to \href{https://aip.scitation.org/journal/phf}{Physics of Fluids.}}
\end{abstract}
\maketitle
\clearpage
\section{Introduction \label{sec:Introduction}}
The smallest insects (body length $<$ 2 mm) such as thrips fly at a chord-based Reynolds number (\rey) on the order of 10, representing what may be considered as the aerodynamic lower limit of flapping flight. Flight at such low \rey~is challenged by significant viscous dissipation of kinetic energy. To overcome viscous losses, tiny insects have to continuously flap their wings to stay aloft. These insects are observed to flap their wings at high frequencies~($\mathcal{O}$(100 Hz)), likely to increase \rey~by increasing their wing tip velocity. In contrast to larger insects such as hawkmoths and fruit flies, tiny insects are also observed to operate their wings at near-maximal stroke amplitudes~\citep{Sane16} and large pitch angles~\citep{XinCheng17,YuZhu19}. At large stroke amplitudes, the wings of tiny insects come together in close proximity of each other at the end of upstroke (`clap') and move away from each other at the start of downstroke (`fling'). Since the discovery of `clap-and-fling' by~Weis-Fogh~\cite{Weis-Fogh73} in the small chalcid wasp \textit{Encarsia Formosa}, this mechanism has been observed in the free flight of other tiny insects such as the greenhouse whitefly~\cite{Weis-Fogh75}, thrips ~\citep{Ellington84kin,Santhanakrishnan14}, parasitoid wasps~\cite{Miller09} and jewel wasps~\cite{Miller09}. A number of studies have explored the fluid dynamics of clap-and-fling experimentally~\citep{Maxworthy79,Spedding86, Lehmann07}, theoretically~\citep{Lighthill73,Ellington84kin,Kolo11}, and numerically~\citep{Miller04, Santhanakrishnan14, Jones15, MaoSun06,Kolo11, Arora14}, and have found that wing-wing interaction augments lift force through the generation of bound circulation at the leading edges of the wings during fling~\citep{Lighthill73,Maxworthy79,Spedding86,Miller05,Kolo11}. 

In contrast to larger flying insects where a stable leading edge vortex (LEV) is observed with a shed trailing edge vortex (TEV)~\cite{Birch04}, previous studies of a single wing in linear translation~\cite{Miller04} and in semi-circular revolution~\cite{Santhanakrishnan18} have shown that lift generation at \rey$\sim$10 is reduced due to `vortical symmetry', where both the LEV and TEV remain attached to the wing. Miller and Peskin~\cite{Miller05} showed that lift enhancement by clap-and-fling is more pronounced for \rey$\sim\mathcal{O}$(10) than at higher \rey, as most of the lift lost during the downstroke and upstroke (on account of vortical symmetry) can be recovered by establishing LEV-TEV vortical asymmetry during wing-wing interaction. However, at \rey~relevant to tiny insect flight, Miller and Peskin~\cite{Miller05} also showed that large drag penalties are associated with the fling. Subsequent studies have since shown that wing flexibility and the unique bristled structure of tiny insect wings can provide aerodynamic benefits by lowering drag forces needed to fling wings apart and increasing lift over drag ratio~\cite{Miller09,Santhanakrishnan14,Jones16,Kasoju18,Ford19}.

Forces generated by biological bristled structures such as tiny insect wings depend on inter-bristle flow that is a function of~\rey~based on bristle diameter ($Re_b$). Previous studies~\citep{Cheer87,Loudon94} have shown that an array of bristles can undergo transition from acting as a leaky rake to a solid paddle with decreasing $Re_b$. Dynamically scaled models of bristled wings during translation and rotation have been reported to show little variation in forces in comparison with a solid wing~\citep{Sunada02,kolomenskiy2020}. Further, studies using comb-like wings~\citep{Weihs08, Davidi12} were found to generate almost the same amount of forces as a solid wing, with a 90\% drop in wing weight. Recent studies using bristled wings~\citep{Lee17, Lee18, Lee20} observed the formation of diffused shear layers around the bristles at smaller inter-bristle gaps. These shear layers prevent fluid from leaking through the inter-bristle gaps, resulting in the bristled wing behaving similar to a solid wing. A central limitation of the above studies is the lack of considering clap-and-fling kinematics observed in freely-flying tiny insects, involving aerodynamic interaction of bristled wing pairs. In our recent study~\cite{Kasoju18} examining clap-and-fling of bristled wing pairs at $Re\sim\mathcal{O}(10)$, we found that leaky flow through the bristles results in large drag reduction and disproportionally lower lift reduction (i.e., improved lift over drag ratio) when compared to forces generated by geometrically equivalent solid wings. These aerodynamic benefits were diminished at $Re$=120 (relevant to larger fruit flies)~\cite{Ford19}, suggesting that the use of clap-and-fling in conjunction with bristled wings is particularly well-suited at \rey~relevant to tiny insect flight.

\begin{figure*}
    \centering
    \includegraphics[width=1\textwidth]{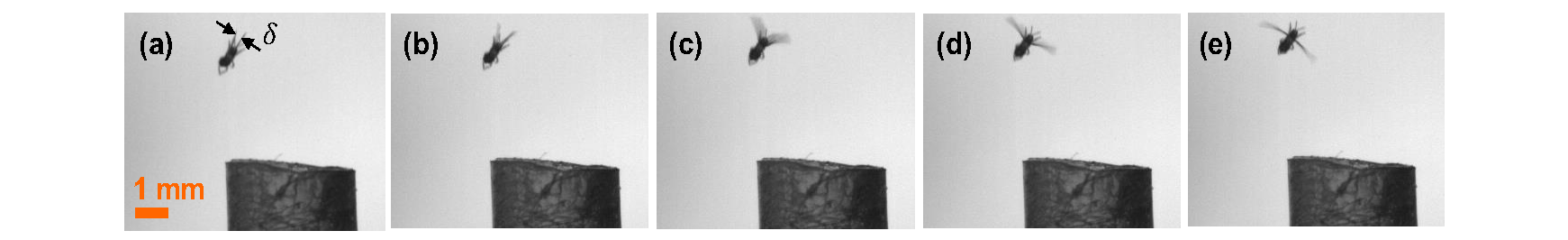}
    \caption{Successive snapshots of thrips in free take-off flight. (a) End of upstroke (`clap') where the wings come in close proximity of each other, separated by non-dimensional inter-wing spacing $\delta$. (b) Start of downstroke (`fling') where the wings move apart from each other, followed by the rest of the downstroke from (c) to (e). $\delta$ ranges from about 10$\%$ to 25$\%$ of the wing chord. 
}
    \label{fig1}
\end{figure*}

In terms of wing-wing interaction of bristled wings, our recordings of free-takeoff flight of thrips show that these insects bring the wings close together ($\sim$1/10-1/4 of chord length) at the end of upstroke (clap) before flinging the wings apart~({\bf Figure~\ref{fig1}}). Previous studies~\citep{MaoSun06, Arora14} have found that increasing initial inter-wing spacing ($\delta$ in {\bf Figure~\ref{fig1}}, expressed non-dimensionally as \% of chord length) of interacting solid wings decreases aerodynamic forces. For $\delta$>80\%, interference effects between the wings were found to diminish. A high pressure region was observed to form between the interacting solid wings during the end of the clap phase that generated a sharp peak in forces at the end of clap and start of fling~\citep{Cheng17}. However, none of these studies examined how inclusion of wing bristles impacts clap-and-fling aerodynamics under varying $\delta$. In terms of wing motion, a recent study reported the wing kinematics of free-flying thrips~\citep{YuZhu19} and noted large changes in pitch angle for small changes in revolution of the wing. While this indicates that thrips wings may purely rotate at the start of fling before translation, it remains unknown as to whether there are aerodynamic benefits associated with such kinematics. In this study, we aimed to examine how varying $\delta$ and wing kinematics impacts aerodynamic interaction of bristled wings during fling at \rey=10. We used a dynamically scaled robotic platform fitted with a pair of physical bristled wing models for investigation. Aerodynamic force measurements and flow visualization were conducted for varying $\delta$ in the range of 10\% to 50\% for three different kinematics:~1) wings purely rotating about their trailing edges;~2) linear translation of each wing at a fixed angle relative to the vertical; and~3) overlapping rotation and translation of each wing.

\section{Methods}\label{sec:methods}
\subsection{Dynamically scaled robotic platform}
\begin{figure*}
	{\centering\includegraphics[width=0.85\textwidth]{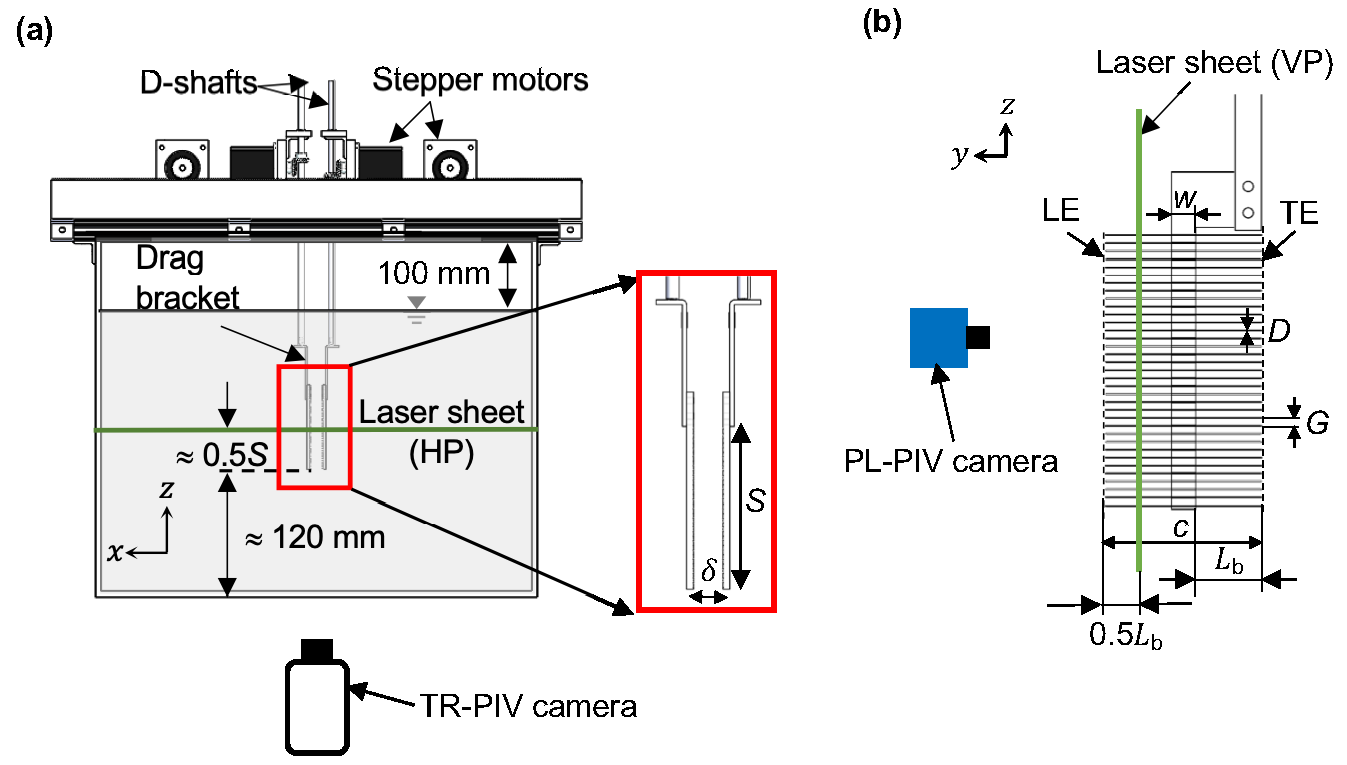}}
		\caption{Robotic platform and experimental setup used for force and PIV measurements. (a) Front view of the robotic platform with a pair of scaled-up physical bristled wing models separated by initial inter-wing spacing $\delta$ expressed non-dimensionally as \% of wing chord ($c$). Tank dimensions were 510 mm (length)$\times$510 mm (width)$\times$410 mm (height). 2D TR-PIV setup with high-speed camera and laser sheet along a horizontal plane (HP) located at a spanwise location of approximately 0.5$S$ was used to visualize the flow generated around the chord of both wings. (b) Magnified view of rectangular bristled wing model used in this study with chord $c$ and span $S$. The bristled wing consisted of a 3 mm thick solid membrane of width $w$ and length $S$, and 35 circular bristles (bristle diameter, $D$=0.2032 mm) were attached on both sides. Inter-bristle gap $G$ was maintained constant throughout the wing at 2.0 mm, such that the ratio $G$/$D$ $\approx$ 10 was in the biological range of tiny insects. 2D PL-PIV measurements were conducted to examine the inter-bristle flow along the wing span, using an sCMOS camera focused on a laser sheet along a vertical plane (VP) located at 50\%$L_\text{b}$ referenced from the LE. LE = leading edge; TE = trailing edge; $x$, $y$, $z$ are fixed coordinate definitions that are used throughout this study; $c$=wing chord=45 mm; $S$=wing span=81 mm; total number of bristles=70; $w$=membrane width=7 mm; $L_\textnormal{b}$=bristle length on each side of the membrane=19 mm.}
	\label{fig2}
\end{figure*}

We comparatively examined the forces and flows generated during the prescribed motion of a pair of bristled wing physical models to those of a single bristled wing. The wing models were driven by a dynamically scaled robotic platform ({\bf Figure~\ref{fig2}(a)}) that has been used in our previous studies~\citep{Kasoju18,Ford19}. The experimental setup consists of a scaled-up bristled wing pair (or a single bristled wing) immersed in a 510 mm (length)$\times$510 mm (width)$\times$410 mm (height) optically clear acrylic tank filled with glycerin. Each wing was attached to a stainless steel D-shaft (diameter=6.35 mm) using custom made L-brackets~\cite{Kasoju18}. Uniaxial strain gauges were mounted on the L-brackets to measure lift and drag forces. Two 2-phase hybrid stepper motors with integrated encoders (ST234E, National Instruments Corporation, Austin, TX, USA) were used to drive the D-shaft to perform rotational and translational motion. Rotational motion was achieved using a bevel gear coupled to a motor and a D-shaft, while translational motion was achieved using a rack-and-pinion mechanism driven by a second motor. All the stepper motors (4 motors needed for a bristled wing pair, 2 motors needed for a single wing) were controlled using a multi-axis controller (PCI-7350, National Instruments Corporation, Austin, TX, USA) via a custom LabVIEW program (National Instruments Corporation, Austin, TX, USA).

\subsection{Bristled wing models}\label{sec:Bristled wing model}
We fabricated a pair of rectangular scaled-up bristled wing models ({\bf Figure~\ref{fig2}(b)}) with wing span ($S$) of 81 mm and chord ($c$) of 45 mm. The bristled wing consisted of a 3 mm thick solid membrane (laser cut from optically clear acrylic) of length equal to $S$ and 7 mm width ($w$), with 35 bristles of equal length ($L_\text{b}$=19 mm) attached on two opposite sides along the length of the membrane (70 bristles in total, in the range of tiny insects~\cite{Kasoju20}). The bristles consisted of 0.2032 mm diameter ($D$) 304 stainless steel wires, each being cut to length $L_b$. The inter-bristle gap ($G$) was maintained at 2 mm throughout the wing, to obtain $G$/$D$=10 in the range of $G$/$D$ of tiny insect wings~\cite{Jones16,Kasoju20}. An equivalent solid wing pair with the same $S$ and $c$ as the bristled wing was also laser cut from optically clear acrylic for comparative measurements.

\subsection{Kinematics}\label{sec:Kinematics}
\begin{figure*}[b]
    \centering
    \includegraphics[width=1\textwidth]{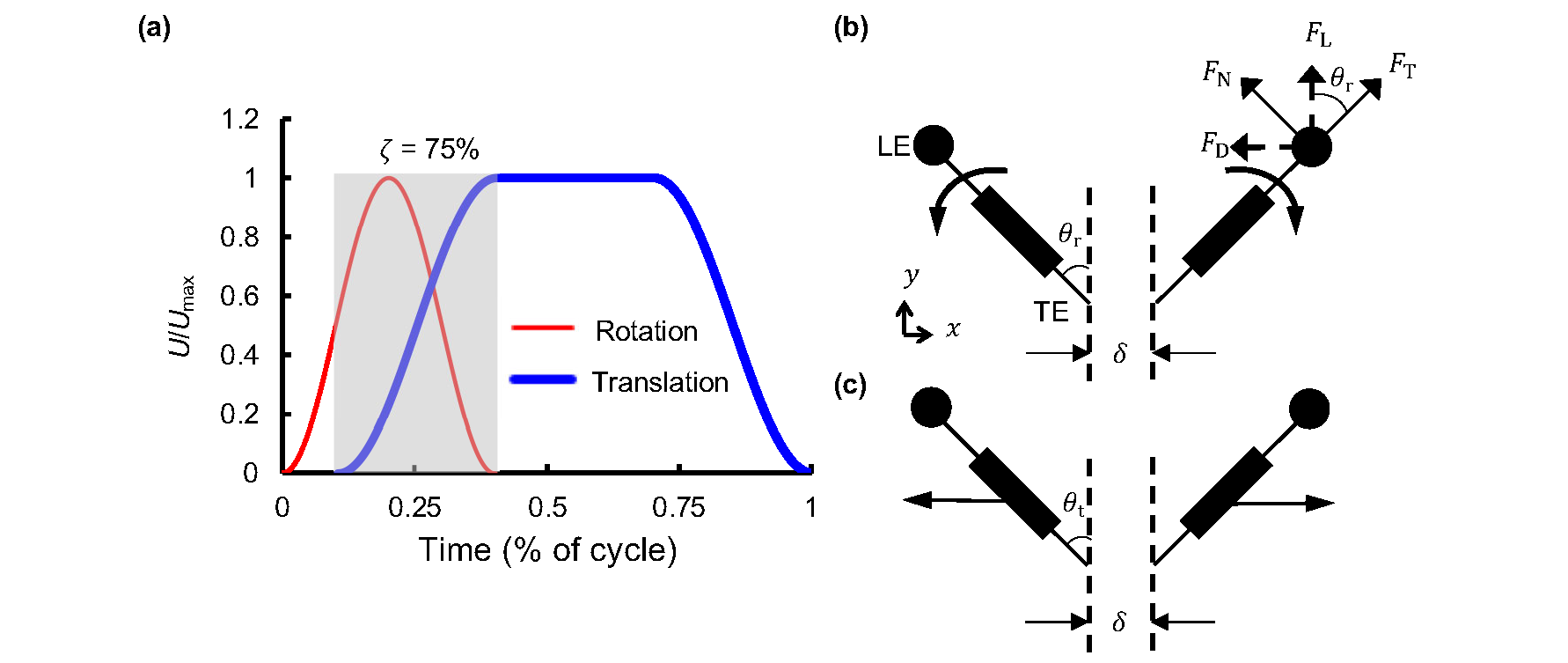}
    \caption{Wing kinematics used in this study. (a) Time-varying motion profile for a single wing based on a previous study by Miller and Peskin~\cite{Miller05}. Instantaneous wing tip velocity $U$ was non-dimensionalized by peak tip velocity $U_{\textnormal{max}}$. $U_{\textnormal{max}}$ was maintained as a constant (0.157 m~s$^{-1}$) throughout the study, and calculated using equation (1) to obtain \rey=10. Time is expressed non-dimensionally in terms of percentage of cycle duration $T$. The value of $T$ was changed for different kinematics (provided in Table~\ref{table1}). Thin and thick lines indicate rotational and translational motion, respectively. $\zeta$ indicates the percentage of overlap between wing rotation and the start of translation, and was varied in this study. (b) and (c) show sectional views of a bristled wing pair during wing rotation and linear translation, respectively. $\theta_{\textnormal{r}}$ is the angle at the end of wing rotation; $\theta_{\textnormal{t}}$ is the translation angle (both $\theta_{\textnormal{r}}$ and $\theta_{\textnormal{r}}$ are defined relative to the vertical). Lift ($F_{\textnormal{L}}$) and drag ($F_{\textnormal{D}}$) forces were separately measured using two custom L-brackets~\cite{Kasoju18}, by taking components of $F_{\textnormal{T}}$ and $F_{\textnormal{N}}$ in the vertical ($F_{\textnormal{L}}$) and horizontal ($F_{\textnormal{D}}$) directions.}
\label{fig3}
\end{figure*}

The robotic platform enabled rotation and linear translation of wing models along a horizontal stroke plane. We examined the isolated and combined roles of rotation and linear translation in this study. Sinusoidal and trapezoidal motion profiles were used for wing rotation and translation, respectively ({\bf Figure~\ref{fig3}(a)}), using equations developed by Miller and Peskin~\cite{Miller05}. The 2D clap-and-fling kinematics developed by Miller and Peskin~\cite{Miller05} has been used in several previous studies~\cite{Miller09, Arora14, Santhanakrishnan14, Jones16, Ford19}. The maximum velocity during rotation and linear translation ($U_\textnormal{max}$) was maintained constant throughout the study at 0.157 m~s$^{-1}$. For tests examining wing rotation, each wing model rotated about its trailing edge (TE) from initial vertical position to an angle $\theta_\textnormal{r}$ relative to the vertical ({\bf Figure~\ref{fig3}(b)}). The peak angular velocity ($\omega_\textnormal{max}$) of wing rotation was maintained constant to obtain the above $U_\textnormal{max}$. The cycle duration ($T$) thus changed with varying $\theta_\textnormal{r}$ (Table~\ref{table1}). For tests examining linear translation, each wing was preset prior to the start of wing motion to a fixed angle ($\theta_\textnormal{t}$) relative to the vertical ({\bf Figure~\ref{fig3}(c)}). 

For tests examining combination of rotation and linear translation, each wing was prescribed to rotate and translate under varying levels of overlap ($\zeta$) that was defined based on the start of wing translation relative to rotation ({\bf Figure~\ref{fig3}(a)}). Note that $\zeta$=0\% means that linear translation started at the end of rotation, and $\zeta$=100\% means that linear translation started at the same time as start of rotation. $\theta_\textnormal{r}$ and $\theta_\textnormal{t}$ of 45$^\circ$ were used for all tests examining combined rotation and linear translation. $\omega_\textnormal{max}$ for each $\zeta$ that was tested was equal to $\omega_\textnormal{max}$ used in tests involving only wing rotation. $T$ varied for each tested condition of combined rotation and linear translation (Table~\ref{table1}). The wing motion for both the wings were identical but opposite in sign. Also, the motion was strictly two-dimensional (2D) without changes in the stroke plane. At the end of every cycle of each test condition, the wings were programmed to move back to the starting position and were paused for at least 30 seconds before starting the next cycle so as to remove the influence of cycle-to-cycle interactions. A description of the mathematical equations used in modeling wing kinematics is provided in the appendix.

\subsection{Test conditions}\label{sec:Test conditions}
Bristled wing pairs and a single bristled wing were tested at \rey=10 for the following kinematics:~1) rotation to $\theta_\textnormal{r}$ values of 22.5$^\circ$, 45$^\circ$, 67.5$^\circ$; 2) linear translation at $\theta_\textnormal{t}$ values of 0$^\circ$ (vertically oriented), 22.5$^\circ$, 45$^\circ$, 67.5$^\circ$; and 3) combined rotation and linear translation for $\zeta$=0\%, 25\%, 50\%, 75\%, 100\%. Each of the above test conditions were repeated for $\delta$=10\%, 30\%, 50\% between the bristled wing pairs as well as in a single bristled wing (latter corresponding to $\delta\rightarrow\infty$). The wing models being tested were fully immersed in 99\% glycerin solution. The kinematic viscosity ($\nu$) of the glycerin used in this study was measured using a Cannon-Fenske routine viscometer (size 400, Cannon Instrument Company, State College, PA, USA) to be 707 mm$^{2}$~s$^{-1}$ at room temperature. To obtain \rey=10, peak velocity ($U_{\textnormal{max}}$) was calculated to be 0.157 m s$^{-1}$ (and maintained constant as mentioned in subsection~\ref{sec:Kinematics}) using the following equation:
\begin{equation}
	Re=\frac{U_{\textnormal{max}} c}{\nu}
	\label{eq:Re}	
\end{equation}
where $c$ ({\bf Figure~\ref{fig2}(b)}) and $\nu$ are constants. Using the kinematics equations provided in Miller and Peskin~\cite{Miller05}, motion profiles were created to drive the stepper motors. $Re$ based on bristle diameter $D$ (defined as $Re_b=U_{\textnormal{max}} D/\nu$) was also maintained constant at 0.045 throughout the study, which is in the range of thrips (0.01-0.07)~\citep{Jones16}. 

\subsection{Force measurements}
Similar to our previous studies~\citep{Kasoju18,Ford19}, force data was collected using L-brackets with uniaxial strain gauges mounted in half-bridge configuration. A strain gauge conditioner continuously measured the forces in the form of voltage signals based on L-bracket deflection during wing motion. Two separate L-brackets were used for non-simultaneous acquisition of lift and drag force data. The design of lift and drag L-brackets and validation of the methodology can be found in Kasoju \etal~\cite{Kasoju18} Lift and drag forces were only measured on one wing in tests involving a bristled wing pair, with the assumption that the forces generated by the other wing would be equal in magnitude (as the motion was symmetric for both wings of a wing pair). The raw voltage data was acquired using a data acquisition board (NI USB-6210, National Instruments Corporation, Austin, TX, USA) once the LabVIEW program (used for driving the motors) triggered to start the recording. Force data and angular position of the wings were acquired during each cycle at a sample rate of 10 kHz for all the test conditions mentioned in subsection~\ref{sec:Test conditions}. The raw data was processed in the same manner as in our previous studies~\citep{Kasoju18,Ford19} and implemented via a custom MATLAB script . A third order low-pass Butterworth filter with a cutoff frequency of 24 Hz was first applied to the raw voltage data. The baseline offset (obtained with wing at rest) was averaged in time and subtracted from the filtered voltage data. The lift and drag brackets were calibrated manually, and the calibration was applied to the filtered voltage data obtained from the previous step to calculate tangential ($F_\textnormal{T}$) and normal ($F_\textnormal{N}$) forces ({\bf Figure~\ref{fig3}(b)}). Lift and drag forces were calculated as components of $F_\text{T}$ and $F_\text{N}$ as described in subsection~\ref{sec:Calculated quantities}.

\subsection{Flow visualization}\label{sec:Flow vis}
 We conducted 2D time-resolved particle image velocimetry (2D TR-PIV) measurements to visualize time-varying chordwise flow generated by the motion of a wing pair (or a single wing) at a horizontal plane (HP) located at mid-span ({\bf Figure~\ref{fig2}(a)}), and quantify the strength of the LEV and TEV. In addition, 2D phase-locked PIV (2D PL-PIV) measurements were conducted to characterize the inter-bristle flow along the wing span at a vertical plane (VP) located at 0.5$L_\text{b}$ measured from the leading edge (LE) as shown in {\bf Figure~\ref{fig2}(b)}. 

\subsubsection{2D TR-PIV}\label{sec:2D TR-PIV}
2D TR-PIV measurements were performed to visualize the flow structures generated by bristled wings in rotation, linear translation and their combination for varying $\delta$.~The glycerine solution was seeded with 55 $\mu$m diameter titanium dioxide filled polyamide particles (density=1.2 g~cm$^{-3}$, LaVision GmbH, G{\"o}ttingen, Germany). Seeding particles were mixed in the glycerin solution at least one day before TR-PIV data acquisition to allow adequate time to realize homogenous initial distribution. The flow field was illuminated using a 527 nm wavelength single cavity Nd:YLF high-speed laser with a maximum repetition rate of 10 kHz and pulse energy of 30 mJ (Photonics Industries International, Ronkonkoma, NY, USA). This laser provided a 0.5 mm diameter beam that was passed through a -20 mm focal length plano-concave cylindrical lens to generate a 3 mm thick laser sheet, which was then oriented horizontally along the mid-span (HP in {\bf Figure~\ref{fig2}A}). Raw TR-PIV images for each of the test conditions were acquired using a high-speed complementary metal-oxide-semiconductor (CMOS) camera (Phantom Miro 110, Vision
Research Inc., Wayne, NJ, USA) with a spatial resolution of 1280$\times$800 pixels, maximum frame rate of 1630 frames~s$^{-1}$, and pixel size of 20$\times$20 microns.~A 50 mm constant focal length lens (Nikon Micro Nikkor, Nikon Corporation, Tokyo, Japan) was attached to the TR-PIV camera with the aperture set to 1.4 for all the measurements. A digital pulse was generated with a LabVIEW program to use as a trigger to begin recording TR-PIV images synchronized to the start of wing motion. TR-PIV recordings were initiated after 10 consecutive cycles to establish a periodic steady-state flow in the tank. For each of the test conditions, 100 images were acquired per cycle for 5 consecutive cycles using frame rates specified in Table~\ref{table1}.

\begin{center}
    \begin{table}[]
	    \begin{tabular}{|c|c|c|}
            \hline
            Kinematics&Cycle duration&Frame rate\\
            &$T$~[ms]&[Hz]\\
            \hline
            Rotation,~$\theta_{\textnormal{r}}$~[$^\circ$]&&\\
            22.5&250&400\\
            45&500&200\\
            67.5&750&133.33\\
            \hline
            Translation,~$\theta_{\textnormal{t}}$ [$^\circ$]&&\\
            0&1110&90\\
            22.5&1110&90\\
            45&1110&90\\
            67.5&1110&90\\
            \hline
            Overlap,~$\zeta$ [\%]&&\\
            0&1610&61.72\\
            25&1490&67.11\\
            50&1360&73.52\\
            75&1240&80.64\\
            100&1110&90.09\\
            \hline
        \end{tabular}
        \caption{Experimental test conditions, cycle duration and TR-PIV frame rates used for: rotation ($\theta_{\textnormal{r}}$), translation ($\theta_{\textnormal{t}}$), and overlapping rotation and translation ($\zeta$ in \%). Note that $\zeta$=0\% indicates translation starts at the end of rotation, and $\zeta$=100\% indicates translation starts at the same time as start of rotation.}
        \label{table1}
    \end{table}
\end{center}
\subsubsection{2D PL-PIV}\label{sec:2D PL-PIV}
2D PL-PIV measurements were performed to examine inter-bristle flow characteristics along the wing span at a plane located at 0.5$L_\text{b}$ measured from the LE (VP in {\bf Figure~\ref{fig2}(b)}). The same seeding particles as those used in TR-PIV were used for PL-PIV measurements. Illumination for PL-PIV measurements was provided using the same laser used for TR-PIV measurements, but in double-pulse mode where two short laser pulses were emitted at a specified pulse separation interval ($dt$). The laser beam was converted into a planar sheet using the same optics as in TR-PIV. $dt$ ranged between 1,500-19,845 $\mu$s across all the test conditions. Raw PL-PIV image pairs separated by $dt$ (frame-straddling mode, 1 image/pulse) were acquired for each of the test conditions using a  scientific CMOS (sCMOS) camera (LaVision GmbH, G\"{o}ttingen, Germany) with a spatial resolution of 2560$\times$2160 pixels and a pixel size of 6.5$\times$6.5 $\mu$m. A 60 mm constant focal length lens (same as the lens used in TR-PIV) was attached to the sCMOS camera with the aperture set to 2.8 for all PL-PIV measurements. The seeding particles illuminated by the laser sheet were focused using this lens. Similar to TR-PIV, a digital trigger signal was generated for PL-PIV using a custom LabVIEW program. This trigger signal was used as a reference to offset PL-PIV image pair acquisition to occur at specific phase-locked time points along the wing motion cycle.

For wing rotation kinematics, 2D PL-PIV data were acquired at 25\%, 37.5\%, 50\%, 62.5\%, 87.5\% of cycle time ($T$) in both bristled and solid wing models (wing pairs separated by $\delta$=10\%, 30\%, 50\% and single wing) for each $\theta_\text{r}$ specified previously in subsection~\ref{sec:Test conditions}. For linear translation kinematics, 2D PL-PIV data were acquired at 16\%, 33\%, 50\%, 66\%, 83\% of cycle time ($T$) in both bristled and solid wing models (wing pairs separated by $\delta$=10\%, 30\%, 50\% and single wing) for each $\theta_\text{r}$ specified previously in subsection~\ref{sec:Test conditions}. For combined rotation and linear translation kinematics, 2D PL-PIV data were acquired at 8 equally-spaced time points within the overlapping and translational portions of the cycle on both the bristled and solid wing models (wing pair at $\delta$=10\% and single wing) for $\zeta$=25\% and 100\%. 5 consecutive cycles of PL-PIV raw image pairs were acquired at every instant in the cycle.
\subsubsection{PIV processing}\label{sec:PIV processing}
Raw TR-PIV image sequences and PL-PIV image pairs were processed in DaVis 8.3.0 software (LaVision GmbH, G\"{o}ttingen, Germany).~Multi-pass cross-correlation was
performed on the raw PIV images with two passes each on an initial window size of 64$\times$64 pixels and a final window size of 32$\times$32 pixels, each with 50$\%$ overlap. Post-processing was performed by rejecting velocity vectors with peak ratio $Q$ less than 1.2. The post-processed 2D velocity vector fields were phase-averaged across 5 consecutive cycles at every time instant where TR-PIV and PL-PIV data were acquired. The phase-averaged 2D velocity vector fields were exported as .DAT files containing: ($x$,$y$,$u$,$v$) from TR-PIV measurements along the $x$-$y$ plane; and ($x$,$z$,$u$,$w$) from PL-PIV measurements along the $x$-$z$ plane. Note that $u$,$v$,$w$ are velocity components along $x$,$y$,$z$ coordinates, respectively. The exported TR-PIV velocity vector fields were further processed to calculate $z$-component of vorticity ($\omega_z$) and pressure distribution. Similarly, the exported PL-PIV velocity vector fields were used to estimate the reverse flow capacity of the bristled wing. Visualization of exported velocity vector fields was performed using Tecplot 360 software (Tecplot, Inc., Bellevue, WA, USA).

\subsection{Definitions of calculated quantities}\label{sec:Calculated quantities}
\subsubsection{Lift and drag coefficients}
Lift force ($F_\text{L}$) and drag force ($F_\text{D}$) were defined along the vertical and horizontal directions, respectively ({\bf Figure~\ref{fig3}(b)}). Dimensionless lift coefficient ($C_\textrm{L}$) and drag coefficient ($C_\textrm{D}$) were calculated using components of measured $F_\text{N}$ and $F_\text{T}$ using the following equations:
\begin{equation}
    C_\textnormal{L}=\frac{F_\text{L}}{0.5\rho{U^{2}_{\textnormal{max}}}A}=\frac{F_{\textnormal{T}}~\cos\theta}{0.5\rho{U^{2}_{\textnormal{max}}}A}
    \label{eq:CL}
\end{equation}
\begin{equation}
    C_\textnormal{D}=\frac{F_\text{D}}{0.5\rho{U^{2}_{\textnormal{max}}}A}=\frac{F_{\textnormal{N}}~\sin\theta}{0.5\rho{U^{2}_{\textnormal{max}}}A}
    \label{eq:CD}
\end{equation}
\noindent where $\theta$ is the instantaneous angular position of the wing relative to the vertical.

\subsubsection{Circulation}
Circulation was calculated to quantify the strength of the LEV and TEV using the $z$-component of vorticity ($\omega_z$). $\omega_z$ was calculated from the exported TR-PIV velocity fields using the following equation implemented in a custom MATLAB script:
\begin{equation}
    \omega_{z} = \frac{\partial{v}}{\partial{x}}-\frac{\partial{u}}{\partial{y}}.
    \label{eq:omegaz}
\end{equation}

Circulation ($\Gamma$) was calculated from $\omega_z$ fields at all time instants and test conditions where TR-PIV data were acquired, using the following equation in a custom MATLAB script:
\begin{equation}
    \Gamma = \iint_S \omega_{z}~ds
    \label{eq:Gamma}
\end{equation}
\noindent where $S$ is the vorticity region for either the LEV or TEV. For a particular kinematics test condition, the maximum absolute values of $\omega_z$ (i.e., $|\omega_z|$) at both LEV and TEV of a bristled wing were identified. A 15\%$|\omega_z|$ high-pass cut-off was next applied to isolate the vortex cores on a single bristled wing performing the same kinematics. The validation for using this cut-off can be found in our previous study.~\citep{Ford19} $\Gamma$ of LEV or TEV was then calculated by selecting a region of interest (ROI) by drawing a box around a vortex core. A custom MATLAB script was used to automate the process of determining the ROI.~\cite{Samaee20} Essentially, we started with a small square box of 2 mm side and compared the $\Gamma$ value with that of a bigger square box of 5 mm side. If the circulation values matched between the 2 boxes, then we stopped further iteration. If the circulation values did not match between the 2 boxes, we increased the size of the smaller box by 3 mm and iterated the process. When calculating $\Gamma$ of a specific vortex (LEV or TEV), we ensured that $\omega_z$ of the oppositely-signed vortex was zeroed out. For example, $\omega_z$ of the negatively-signed TEV was zeroed out when calculating the $\Gamma$ of the positively-signed LEV on the right wing of a wing pair in fling. This allowed us to work with one particular vortex at a time and avoids contamination of the $\Gamma$ estimation, if the box were to overlap with the region of the oppositely-signed vortex. Note that $\Gamma$ in this study is presented for the left wing only, with the assumption that circulation of LEV and TEV generated around the right wing will be equivalent in magnitude but oppositely signed. $\Gamma$ at the LEV and TEV for all the test conditions were negative and positive, respectively, for the left wing. 

\subsubsection{Downwash velocity}
Downwash velocity ($\overline{V_\textrm{y}}$) was defined as the spatially-averaged velocity of the flow deflected downward by the motion of a bristled wing pair. $\overline{V_y}$ calculated using the following equation from TR-PIV velocity vector fields:
\begin{equation}
    \overline{V_{y}} = \frac{1}{N}\left[\sum_{\textnormal{FOV}}~v(x,y)\right]
    \label{eq:AvgDownwashVelocity}
\end{equation}
\noindent where $v(x,y)$ is the vertical component of velocity and $N$ is the total number of grid points within the TR-PIV field of view (FOV).

\subsubsection{Pressure distribution and average pressure coefficient}
Using the algorithm developed by Dabiri \etal\cite{Dabiri14}, unsteady pressure ($p$) distribution was estimated from TR-PIV velocity vector fields. The estimated pressure distribution was visualized in Tecplot software. In addition, we also calculated the spatially-averaged positive and negative pressures across the entire TR-PIV FOV at every time instant using the following equations:
\begin{equation}
    \overline{p_+} = \frac{1}{N_{+}}\left[\sum_{\textnormal{FOV}}~p_{+}(x,y)\right]
    \label{eq:AvgPosPressure}
\end{equation}
\begin{equation}
    \overline{p_-} = \frac{1}{N_{-}}\left[\sum_{\textnormal{FOV}}~p_{-}(x,y)\right]
    \label{eq:AvgNegPressure}
\end{equation}
\noindent where $\overline{p_{+}}$ and $\overline{p_{-}}$ are the average positive pressure and average negative pressure, respectively, estimated in the entire TR-PIV FOV at a particular timepoint. $N_{+}$ and $N_{-}$ are the total number of grid points in ($x$,$y$) of the portion of the FOV containing positive and negative pressures, respectively. Similar to $\Gamma$ calculation, a cutoff of 15$\%$ of the maximum pressure magnitude of a single bristle wing was applied in these calculations.

Using the average positive and negative pressures, an average coefficient of pressure ($\overline{C_{p}}$) was calculated using the following equation:
\begin{equation}
\overline{C_{p}} = \frac{2~\overline{p}}{\rho U_{\text{max}}^2}
\label{eq:CP}
\end{equation}
\noindent where $\overline{p}$ is the average negative or positive pressure calculated from equations~(\ref{eq:AvgPosPressure}),~(\ref{eq:AvgNegPressure}), and $\rho$ is the density of the fluid medium ($\rho$ of the glycerin solution was measured to be 1259 kg m$^{-3}$).

\subsubsection{Reverse flow capacity (RFC)}
Inter-bristle flow along the wing span is influenced by $Re_b$, $G$, $D$ and wing inclination relative to the flow. Significant changes can be expected in the $Re_b$ range of tiny insect flight, such that the wing bristles can permit fluid leakage or  behave like a solid plate. From the PL-PIV velocity fields, we estimated the capacity of a bristled wing to leak flow (in the direction opposite to wing motion) by comparing the volumetric flow rate (per unit width) along the wing span to that of a geometrically equivalent solid wing undergoing the same wing motion. Reverse flow capacity (RFC) was calculated along a line `L' parallel to the span and located at a distance of $\sim$50\%$L_\text{b}$ ({\bf Figure~\ref{fig2}(b)}). Volumetric flow rate per unit width for a particular wing model ($Q_\text{wing}$) was calculated using the following equation:
\begin{equation}
    Q_{\textnormal{wing}}=\int_L{u~dz}
    \label{eq:qwing}
\end{equation}
\noindent where $u$ denotes the horizontal component of velocity along line `L'. RFC was calculated using the following equation:
\begin{equation}
    \textnormal{RFC}~[\%]=\frac{Q_\text{solid}-Q_\text{bristled}}{Q_\text{solid}} 
    \label{eq:FlowReduction}
\end{equation}
\noindent where $Q_\text{solid}$ and $Q_\text{bristled}$ represents the volumetric flow rate per unit width displaced by a solid wing and bristled wing undergoing the same motion, respectively.

\section{Results}\label{sec: results}
\subsection{Bristled wings in rotation}
\begin{figure*}
    \centering
    \includegraphics[width=0.8\textwidth]{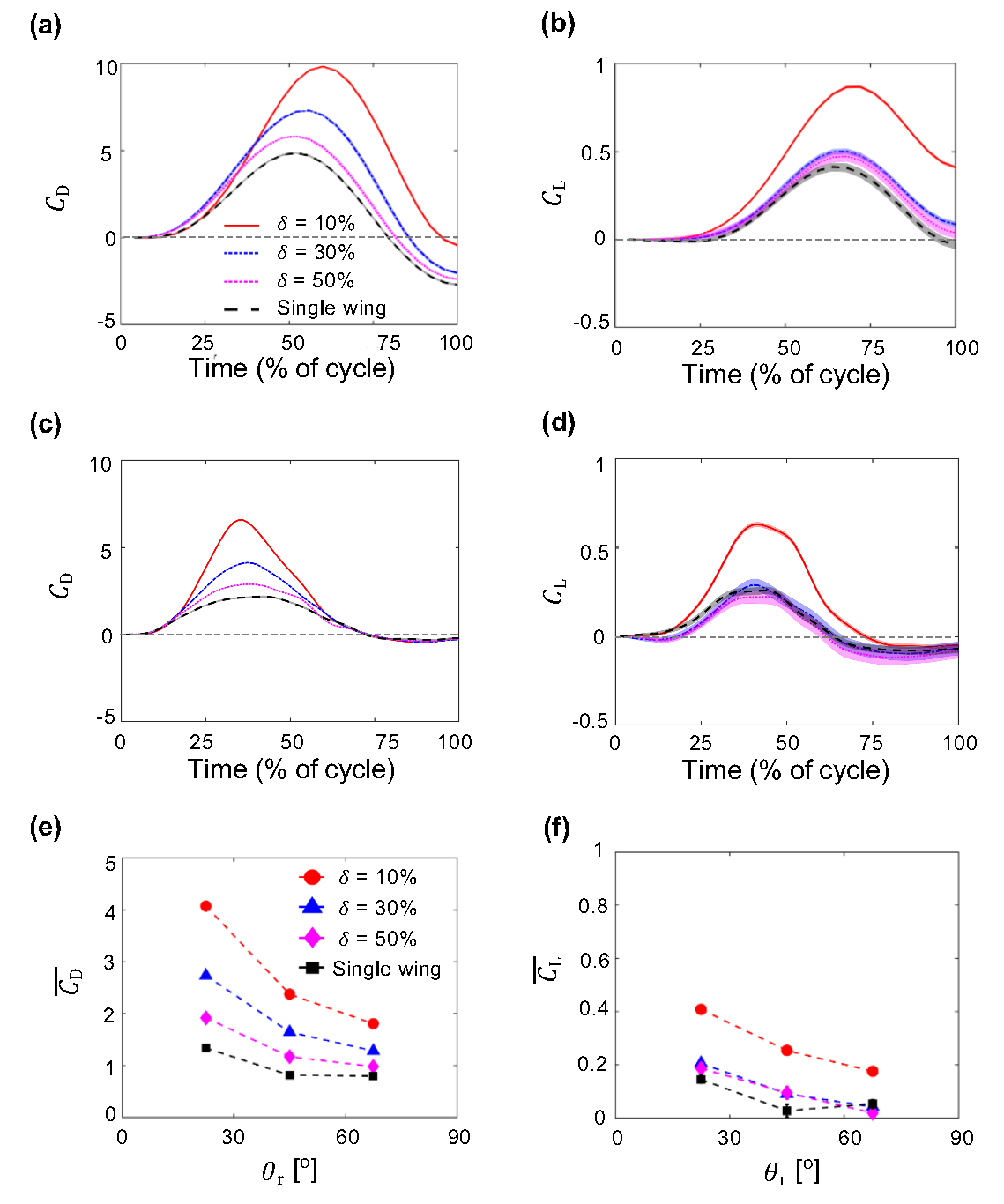}
    \caption{Force coefficients during bristled wing rotation at \rey=10. Shading around each curve represents $\pm$1 standard deviation (SD) across 30 cycles.~(a) and (b) show time-variation of drag coefficient ($C_\textnormal{D}$) and lift coefficient ($C_\textnormal{L}$), respectively, for $\theta_\textnormal{r}$=22.5$^\circ$.~(c) and (d) show time-variation of $C_\textnormal{D}$ and $C_\textnormal{L}$, respectively, for $\theta_\textnormal{r}$=67.5$^\circ$.~(e) and (f) show cycle-averaged drag coefficient ($\overline{C_\textnormal{D}}$) and cycle-averaged lift coefficient ($\overline{C_\textnormal{L}}$), respectively, for varying $\theta_\textnormal{r}$.~Legend for (b)-(d) is shown in (a); legend for (f) is shown in (e).}
    \label{fig4}
\end{figure*}
\noindent\underline{\bf\em Aerodynamic force generation.} In general, both $C_\text{L}$ and $C_\text{D}$ followed the kinematic profile of rotational motion ({\bf Figure~\ref{fig4}(a)-(d)}). When $\theta_\text{r}$ was increased from 22.5$^\circ$ to 67.5$^\circ$, $C_\text{D}$ and $C_\text{L}$ peaks occurred earlier in time ({\bf Figure~\ref{fig4}(c),(d)}).~With increasing $\theta_\text{r}$, disproportionally larger reduction in $C_\text{D}$ was observed when compared to $C_\text{L}$ reduction.
~A noticeable drop in $C_\text{D}$ was observed with increasing $\delta$ for all $\theta_\text{r}$. 
 $C_\text{L}$ was highest for the lowest initial inter-wing spacing ($\delta$=10\%) in both $\theta_\text{r}$=22.5$^\circ$ ({\bf Figure~\ref{fig4}(b)}) and $\theta_\text{r}$=67.5 $^\circ$~({\bf Figure~\ref{fig4}(d)}). Increasing $\delta$ from 10\% to 30\% resulted in a noticeable drop in $C_\text{L}$, following which $C_\text{L}$ showed minimal variation for $\delta$=50\% as well as the single wing ({\bf Figure~\ref{fig4}(b),(d)}). This insensitivity of $C_L$ for $\delta$$\geq$30\% was in sharp contrast to $C_\text{D}$ variation with $\delta$~({\bf Figure~\ref{fig4}(a),(c)}). $C_\text{D}$ dropped below zero toward the end of the cycle for $\theta_\text{r}$=22.5$^\circ$~({\bf Figure~\ref{fig4}(a)}), likely due to wing deceleration altering flow around the bristled wing model in a short time span. With increase in $\theta_\text{r}$ to 67.5$^\circ$, the magnitude of negative drag was decreased ({\bf Figure~\ref{fig4}(c)}).

Cycle-averaged drag coefficient ($\overline{C_\text{D}}$) decreased with increasing $\theta_\text{r}$~({\bf Figure~\ref{fig4}(e)}). Increasing $\theta_\text{r}$ from 22.5$^\circ$ to 67.5$^\circ$ for the single wing showed little to no variation in $\overline{C_\text{D}}$. By contrast, the bristled wing pair with lowest $\delta$ (=10\%) showed substantial decrease in $\overline{C_\text{D}}$ with increasing $\theta_\text{r}$. With further increase in $\delta$, $\overline{C_\text{D}}$ decreased with $\theta_\text{r}$ and approached single wing values. Similar to $\overline{C_\text{D}}$, $\overline{C_\text{L}}$ also decreased with increasing $\theta_\text{r}$. Increasing $\delta$ beyond 10\% resulted in little to no variaton in $\overline{C_\text{L}}$. Finally, a disproportionally larger reduction in $\overline{C_\text{D}}$ was observed compared to smaller reduction in $\overline{C_\text{L}}$ across all $\theta_\text{r}$ and $\delta$.

\begin{figure*}
    \centering
    \includegraphics[width=0.8\textwidth]{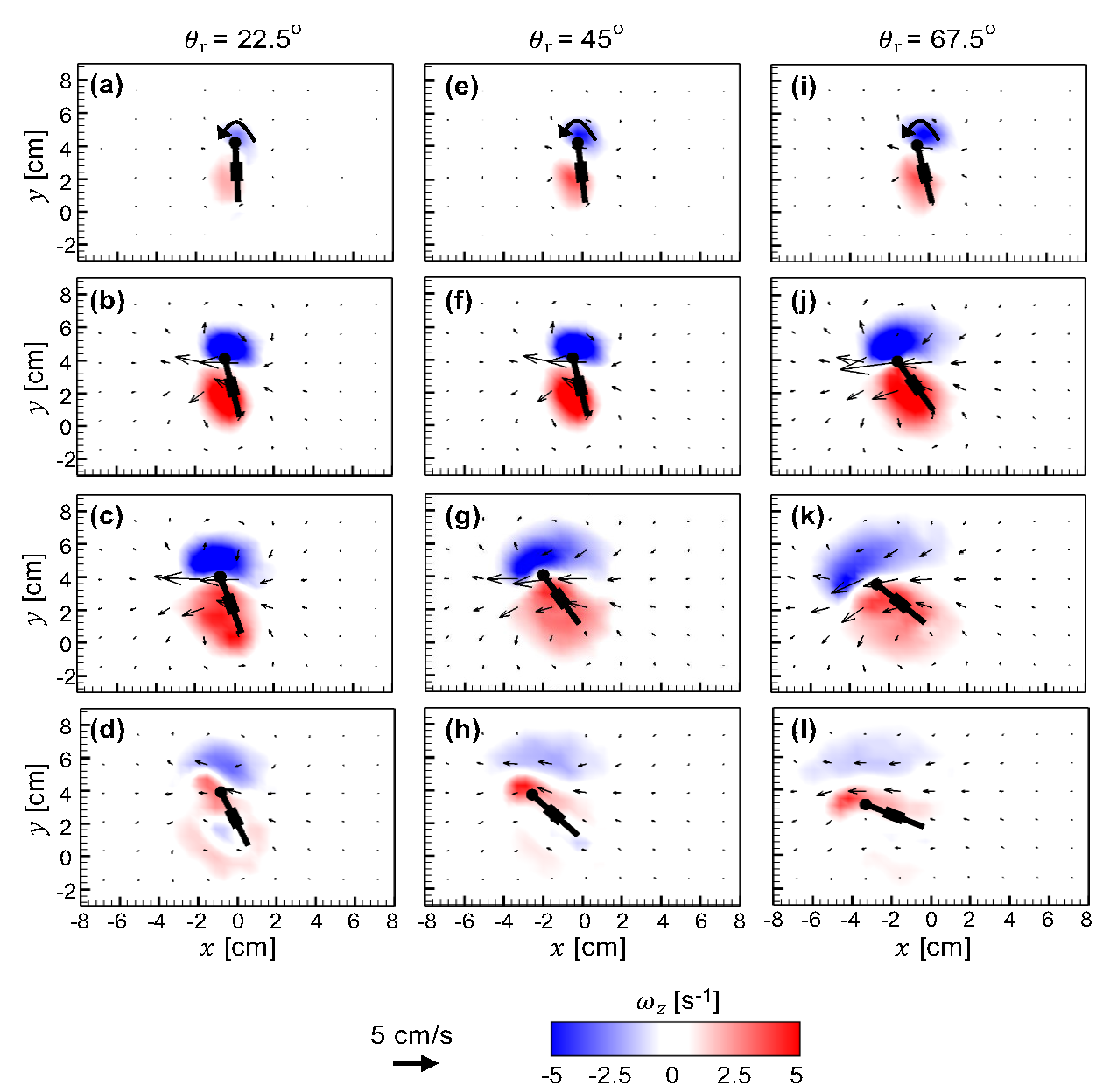}
    \caption{Velocity vectors overlaid on out-of-plane $z$-vorticity ($\omega_z$) contours for a single bristled wing in rotation at \rey=10.~(a)-(d)~$\theta_\textnormal{r}$=22.5$^\circ$;~(e)-(h)~$\theta_\textnormal{r}$=45$^\circ$;~(i)-(l)~$\theta_\textnormal{r}$=67.5$^\circ$. For each $\theta_\textnormal{r}$, 4 timepoints (25\%, 50\%, 75\% and 100\% of cycle time) are shown along each column ((a)-(d); (e)-(h); (i)-(l)) from top to bottom.
}
\label{fig5}
\end{figure*}

\noindent\underline{\bf\em Chordwise flow.} Rotation of a single bristled wing generated a pair of counter-rotating vortices at the LE and TE ({\bf Figure~\ref{fig5}}). For the three $\theta_\text{r}$ values that we examined, we observed both the LEV and TEV to be attached to the wing. Increasing $\theta_\text{r}$ promoted earlier development of the LEV and TEV (compare {\bf Figure~\ref{fig5}(a),(e),(i)}). At 50$\%$ ({\bf Figure~\ref{fig5}(b),(f),(j)}) and 75$\%$ of the cycle ({\bf Figure~\ref{fig5}(c),(g),(k)}), increasing $\theta_\text{r}$ was found to diffuse the vorticity in both the LEV and TEV cores and dissipating at the end of the cycle ({\bf Figure~\ref{fig5}(d),(h),(l)}).

\begin{figure*}[b]
    \centering
    \includegraphics[width=0.88\textwidth]{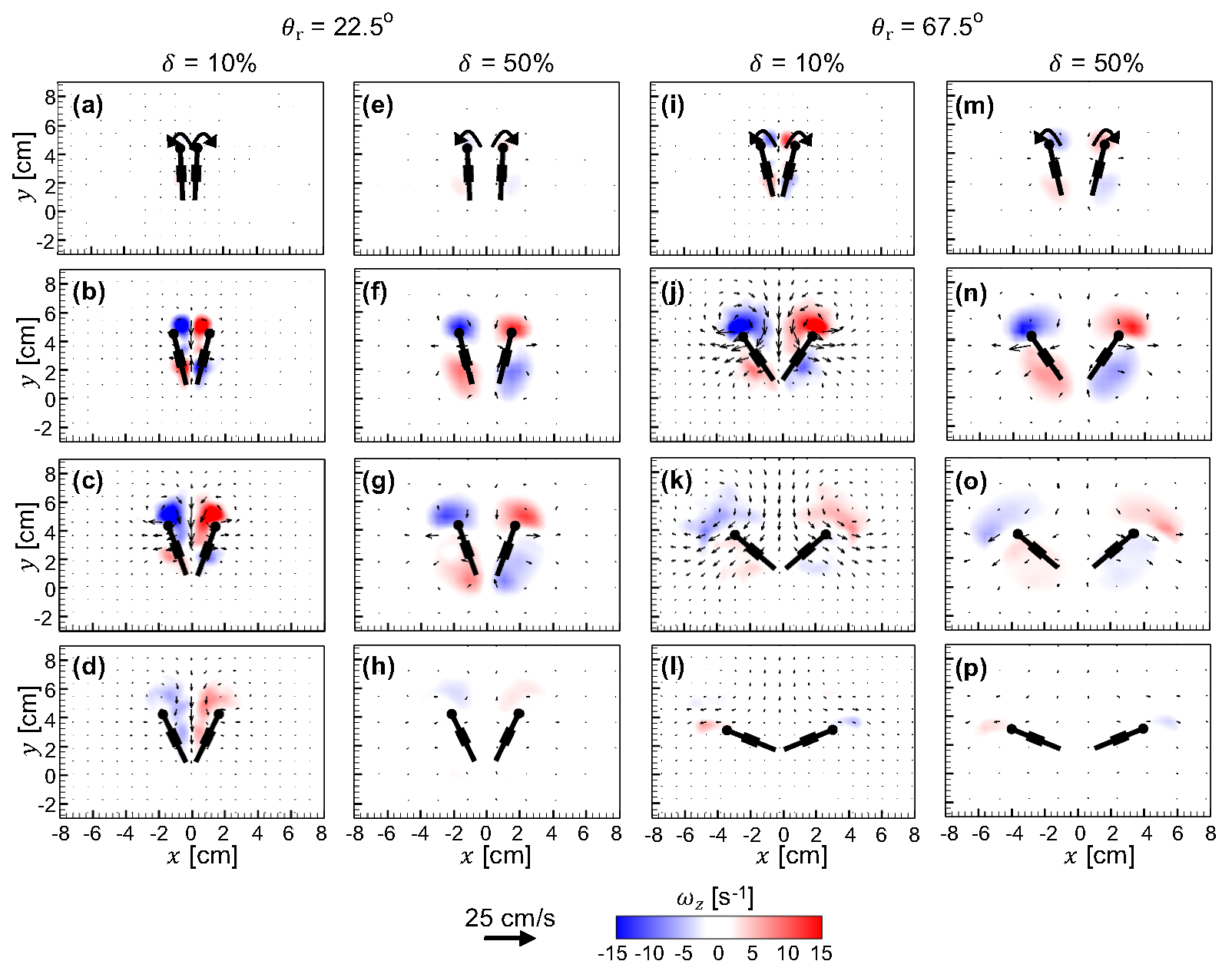}
    \caption{Velocity vectors overlaid on out-of-plane $z$-vorticity ($\omega_z$) contours for a bristled wing pair in rotation at \rey=10.~$\theta_\textnormal{r}$=22.5$^\circ$ is shown for~$\delta$=10\% in (a)-(d) and for~$\delta$=50\% in (e)-(h).~$\theta_\textnormal{r}$=67.5$^\circ$ is shown for~$\delta$=10\% in (i)-(l) and for~$\delta$=50\% in (m)-(p).~For each $\theta_\textnormal{r}$, 4 timepoints (25\%, 50\%, 75\% and 100\% of cycle time) are shown along each column ((a)-(d); (e)-(h); (i)-(l); (m)-(p)) from top to bottom.
}
\label{fig6}
\end{figure*}

For a bristled wing pair that was rotated to $\theta_\text{r}$=22.5$^\circ$, increasing $\delta$ from 10\% ({\bf Figure~\ref{fig6}(a)-(d)}) to 50\% ({\bf Figure~\ref{fig6}(e)-(h)}) diffused the vorticity in both the LEV and TEV. Relative to the LEV for each $\delta$, we observed a weaker TEV (i.e., smaller $\omega_z$) for $\delta$=10\% as compared to $\delta$=50\% ({\bf Figure~\ref{fig6}(a)-(d)}) . The LEV of the bristled wing pair was stronger and smaller in size for smaller $\delta$ compared to the LEV of bristled wing with larger $\delta$ ({\bf Figure~\ref{fig6}(e)-(h)}) that was weaker and more diffused. Similar to the single wing, LEV and TEV of the bristled wing pair for both $\delta$=10\% and 50\% was found to increase in size with increasing cycle duration ($T$) before dissipating at the end of the cycle (100\%$T$).  

Similar to the observations at $\theta_\text{r}$=22.5$^\circ$, increasing $\delta$ diffused and decreased the strength of both the LEV and TEV when the bristled wing pair was rotated to $\theta_\text{r}$ = 67.5$^\circ$ (compare {\bf Figure~\ref{fig6}(i)-(l)} and {\bf Figure~\ref{fig6}(m)-(p)}). In contrast to $\theta_\text{r}$ = 22.5$^\circ$ where LEV and TEV were found to increase in strength from 50\%$T$ to 75\%$T$ ({\bf Figure~\ref{fig6}(b),(c)}), we observed a drop in strength of both the LEV and TEV for $\theta_\text{r}$ = 67.5$^\circ$ for both $\delta$=10\% and 50\% ({\bf Figure~\ref{fig6}(j),(k)}).

\begin{figure*}[b]
    \centering
    \includegraphics[width=0.8\textwidth]{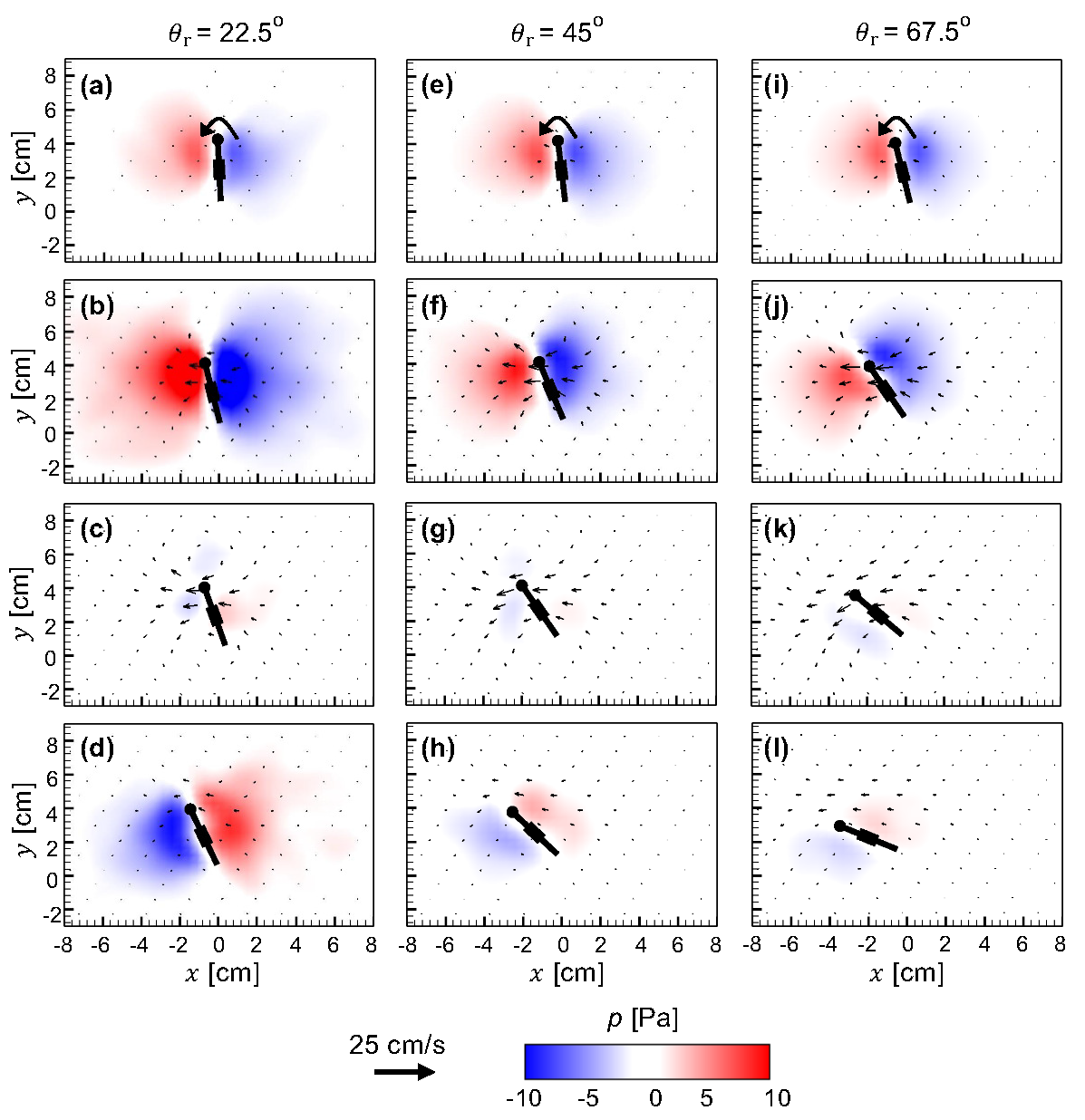}
    \caption{Velocity vectors overlaid on pressure ($p$) contours for a single bristled wing in rotation at \rey=10.~(a)-(d)~$\theta_\textnormal{r}$=22.5$^\circ$;~(e)-(h) ~$\theta_\textnormal{r}$=45$^\circ$;~(i)-(l)~$\theta_\textnormal{r}$=67.5$^\circ$.~For each $\theta_\textnormal{r}$, 4 timepoints (25\%, 50\%, 75\% and 100\% of cycle time) are shown along each column ((a)-(d); (e)-(h); (i)-(l)) from top to bottom. Pressure distribution was calculated from measured velocity fields using the algorithm developed by Dabiri \etal\cite{Dabiri14}
}
\label{fig7}
\end{figure*}

\noindent\underline{\bf\em Pressure distribution.} Positive and negative pressure regions were observed below (i.e., front surface of the wing that first encounters fluid during rotation) and above (back surface of the wing) the single bristled wing in rotation, respectively ({\bf Figure~\ref{fig7}}). Time-variation of pressure distribution around the single rotating wing was similar for all $\theta_\text{r}$ conditions (22.5$^\circ$,45$^\circ$,67.5$^\circ$). Interestingly, we observed the pressure distribution in all $\theta_\text{r}$ conditions to approach zero at 75\%$T$ ({\bf Figure~\ref{fig7}(c),(g),(k)}), which corresponds to right after the start of wing deceleration. In addition, the pressure distribution around the wing flipped in sign at the end of the rotation (100\%$T$;~{\bf Figure~\ref{fig7}(d),(h),(l)}), so that the positive pressure region was located above the wing and negative pressure region was located below the wing. This pressure reversal was particularly pronounced for the smallest $\theta_\text{r}$=22.5$^\circ$ ({\bf Figure~\ref{fig7}(d)}). At 50\%$T$, we observed the pressure distribution to be more diffused for the smallest $\theta_\text{r}$=22.5$^\circ$ ({\bf Figure~\ref{fig7}(b)}) as compared to $\theta_\text{r}$=67.5$^\circ$ ({\bf Figure~\ref{fig7}(j)}).

\begin{figure*}[b]
    \centering
    \includegraphics[width=0.88\textwidth]{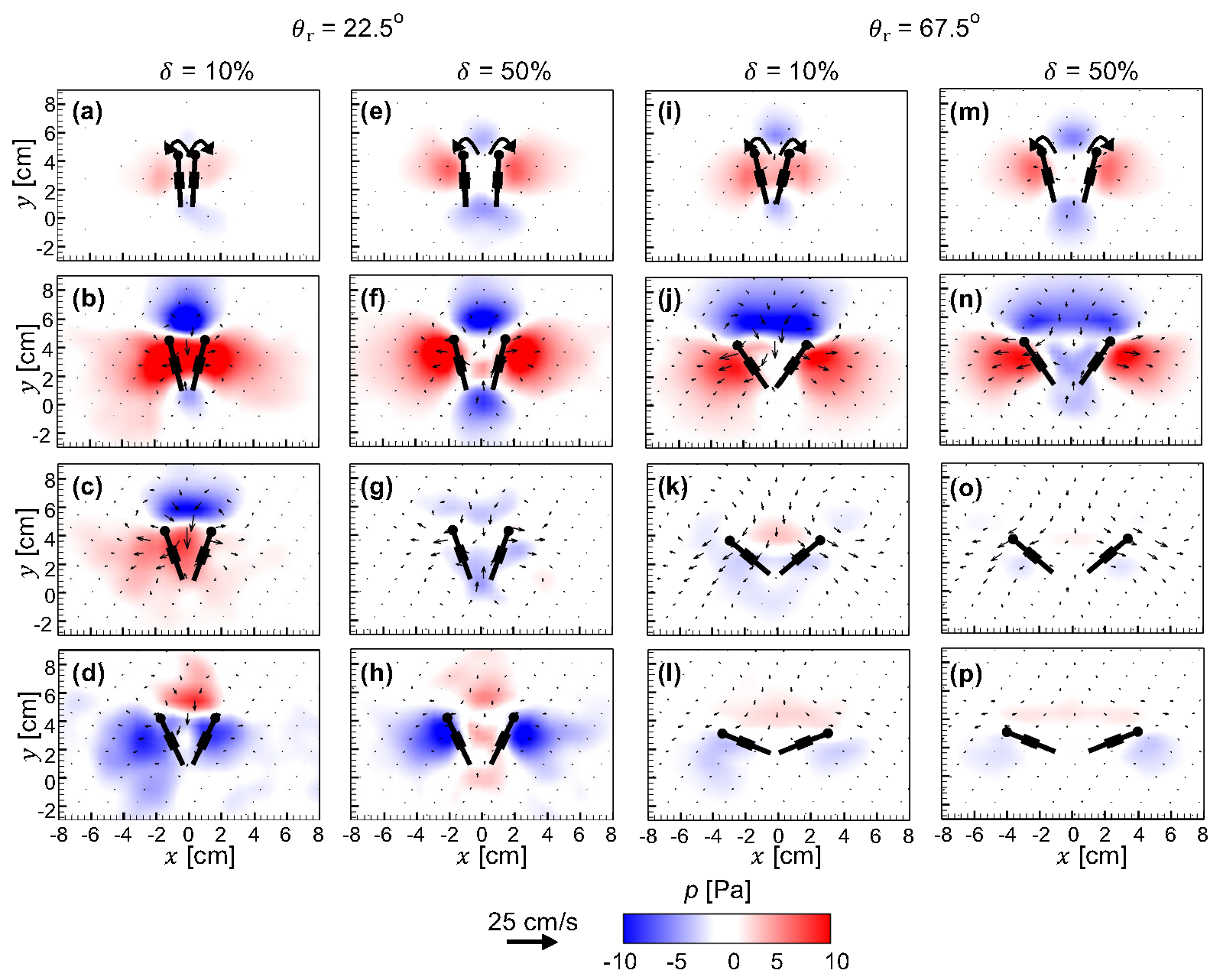}
    \caption{Velocity vectors overlaid on pressure ($p$) contours for a bristled wing pair in rotation at \rey=10.~$\theta_\textnormal{r}$=22.5$^\circ$ is shown for~$\delta$=10\% in (a)-(d) and for~$\delta$=50\% in (e)-(h).~$\theta_\textnormal{r}$=67.5$^\circ$ is shown for~$\delta$=10\% in (i)-(l) and for~$\delta$=50\% in (m)-(p).~For each $\theta_\textnormal{r}$, 4 timepoints (25\%, 50\%, 75\% and 100\% of cycle time) are shown along each column ((a)-(d); (e)-(h); (i)-(l); (m)-(p)) from top to bottom.
}
\label{fig8}
\end{figure*}

Pressure distribution around a bristle wing pair in rotation ({\bf Figure~\ref{fig8}}) was found to be completely different as compared to that of a rotating single wing ({\bf Figure~\ref{fig7}}). During the initial stages of rotational motion, a diffused negative pressure region was observed near the LEs, just above the `cavity' (i.e., inter-wing space) between the two wings ({\bf Figure~\ref{fig8}(a),(e),(i),(m)}).~A weaker negative pressure region was also observed near the TEs, just below the cavity between the two wings. In addition, a diffused region of positive pressure was observed below each wing. For $\delta$ = 10$\%$ and $\theta_\text{r}$=22.5$^\circ$, we observed a diffused region of positive pressure to be distributed in the cavity between the wing pair at 50\%$T$ ({\bf Figure~\ref{fig8}(b)}). The magnitude of positive pressure in the cavity decreased with increasing cycle time. Similar to the single wing model, we observed the positive and negative pressure regions to flip positions at the end of the cycle (100\%$T$; {\bf Figure~\ref{fig8}(d),(h),(l),(p)}). Increasing $\delta$ to 50\% reduced the positive pressure between the wings and simultaneously increased the magnitude of negative pressure near the TEs (compare {\bf Figure~\ref{fig8}(b)} and {\bf Figure~\ref{fig8}(f)}). At 75\%$T$ for  $\theta_\text{r}$=22.5$^\circ$ and $\delta$=10\% ({\bf Figure~\ref{fig8}(c)}), we found both the positive and negative pressure distribution around the wings to substantially decrease in strength. 

Time-variation of pressure distribution around a bristled wing pair rotated to $\theta_\text{r}$=67.5$^\circ$ resembled that of $\theta_\text{r}$=22.5$^\circ$. However, the positive pressure region in the cavity between the wings for $\delta$=10\% and $\theta_\text{r}$=22.5$^\circ$ ({\bf Figure~\ref{fig8}(b)}) was essentially absent for $\delta$=10\% and $\theta_\text{r}$=67.5$^\circ$ ({\bf Figure~\ref{fig8}(j)}). Increasing $\theta_\text{r}$ to 67.5$^\circ$ allowed the negative pressure region near the LEs (above the cavity) to diffuse over a larger region as compared to $\theta_\text{r}$=22.5$^\circ$. In contrast to increasing $\delta$ for $\theta_\text{r}$=22.5$^\circ$ ({\bf Figure~\ref{fig8}(f)}), increasing $\delta$ for $\theta_\text{r}$=67.5$^\circ$ resulted in negative pressure distribution in the cavity between the wing at 50$\%$ cycle time ({\bf Figure~\ref{fig8}(n)}).

\subsection{Bristled wings in linear translation}
\begin{figure*}[b]
    \centering
    \includegraphics[width=0.8\textwidth]{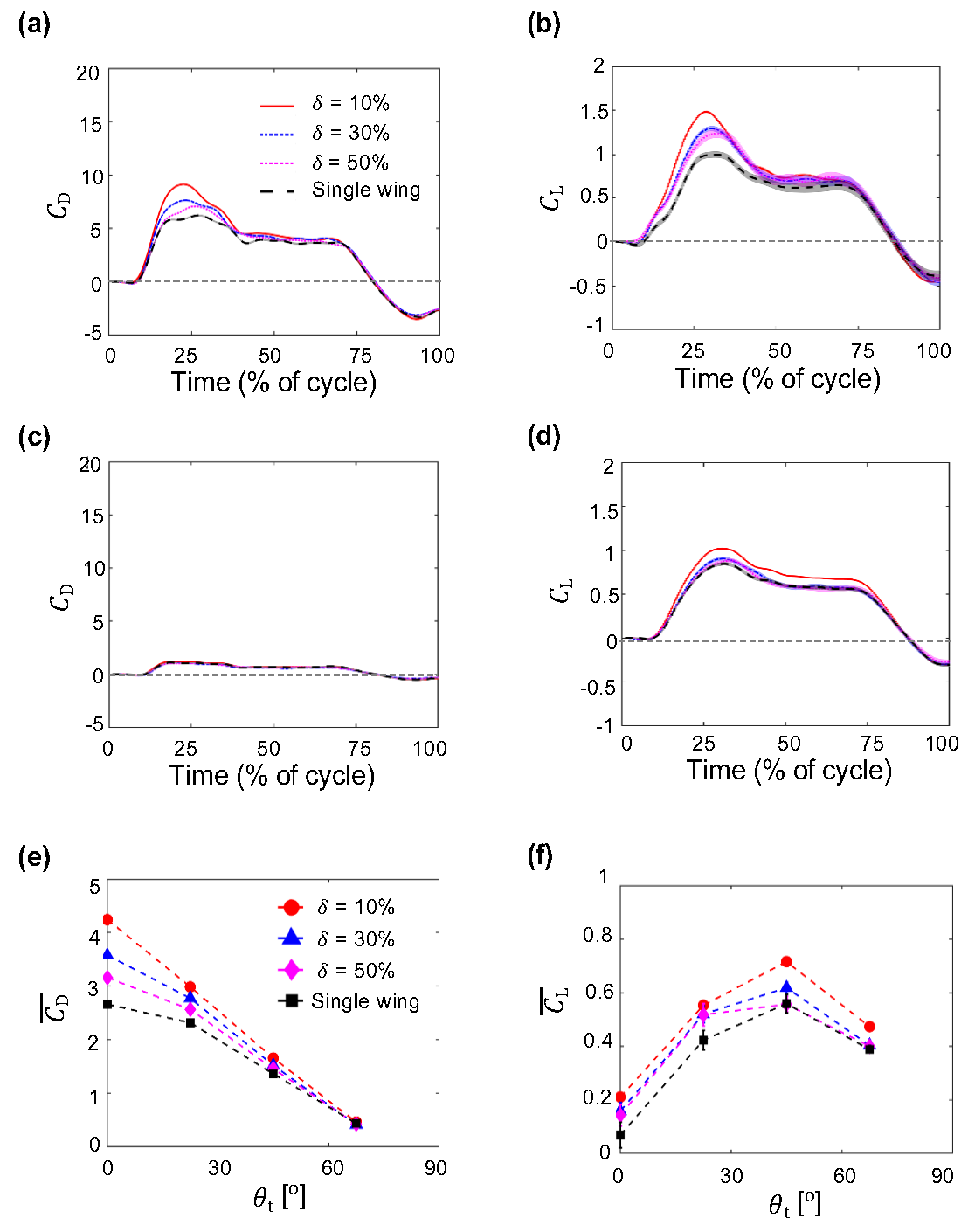}
    \caption{Force coefficients during linear translation of bristled wings at \rey=10. Shading around each curve represents $\pm$1 SD across 30 cycles.~(a) and (b) show time-variation of $C_\textnormal{D}$ and $C_\textnormal{L}$, respectively, for $\theta_\textnormal{t}$=22.5$^\circ$.~(c) and (d) show time-variation of $C_\textnormal{D}$ and $C_\textnormal{L}$, respectively, for $\theta_\textnormal{t}$=67.5$^\circ$.~(e) and (f) show cycle-averaged coefficients $\overline{C_\textnormal{D}}$ and $\overline{C_\textnormal{L}}$, respectively, for varying $\theta_\textnormal{t}$.~Legend for (b)-(d) is shown in (a); legend for (f) is shown in (e).}
\label{fig9}
\end{figure*}

\noindent\underline{\bf\em Aerodynamic force generation.} In general, both $C_\text{L}$ and $C_\text{D}$ were observed to follow similar trends throughout a cycle ({\bf Figure~\ref{fig9}}). For all translational angles ($\theta_\text{t}$) that were tested, we observed an increase in $C_\text{L}$ and $C_\text{D}$ during translational acceleration (see {\bf Figure~\ref{fig3}(a)} for prescribed translation motion profile), followed by $C_\text{L}$ and $C_\text{D}$ remaining approximately constant during constant velocity translation, and a subsequent drop in $C_\text{L}$ and $C_\text{D}$ during translational deceleration ({\bf Figure~\ref{fig9}(a),(b)}). When $\theta_\text{t}$ was increased from 22.5$^\circ$ to 67.5$^\circ$, we observed a disproportionally large reduction in $C_\text{D}$ compared to reduction in $C_\text{L}$ (compare ({\bf Figure~\ref{fig9}(a),(b)}) and {\bf Figure~\ref{fig9}(c),(d)}). In addition, increasing $\theta_\text{t}$ decreased peak values of $C_\text{L}$ and $C_\text{D}$ during translational acceleration by a larger extent as compared to reduction in peak coefficients during constant velocity translation. Similar to wing rotation, we observed $C_\text{D}$ and $C_\text{L}$ to drop below zero toward the end of the cycle for $\theta$=22.5$^\circ$ ({\bf Figure~\ref{fig9}(a),(b)}). A noticeable drop in $C_\text{D}$ and $C_\text{L}$ was observed with increasing $\delta$ for $\theta_\text{t}$=22.5$^\circ$. Increasing $\theta_\text{t}$ to 67.5$^\circ$ decreased the drop in $C_\text{D}$ and $C_\text{L}$ that was observed with increasing $\delta$. Interestingly, changing $\delta$ was found to affect $C_\text{D}$ and $C_\text{L}$ mostly during translational acceleration when the wings were closer to each other, promoting wing-wing interaction. After translational acceleration, when the wings translated further apart, $C_\text{L}$ and $C_\text{D}$ of the bristled wing pair for all $\delta$ values were similar to those generated by a single translating wing.

$\overline{C_\text{D}}$ decreased with increasing $\theta_\text{t}$, and increasing $\delta$ also resulted in decreasing $\overline{C_\text{D}}$ for lower values of $\theta_\text{t}$.~$\overline{C_\text{D}}$ was mostly independent of $\delta$ for $\theta_\text{t}\geq$45$^\circ$, suggesting that increasing $\theta_\text{t}$ reduces wing-wing interaction. In sharp contrast to 
$\overline{C_\text{D}}$, $\overline{C_\text{L}}$ increased with increasing $\theta_\text{t}$ until 45$^\circ$ and subsequently decreased for $\theta_\text{t}$=67.5$^\circ$ ({\bf Figure~\ref{fig9}(f)}). This suggests substantial changes in flow field likely occur for $45^
\circ<\theta_\text{t}\leq67.5^\circ$ to reduce $\overline{C_\text{L}}$ in this range. Increasing $\delta$ resulted in lower variation of $\overline{C_\text{L}}$ as compared to $\overline{C_\text{D}}$. 

\begin{figure*}[b]
    \centering
    \includegraphics[width=0.88\textwidth]{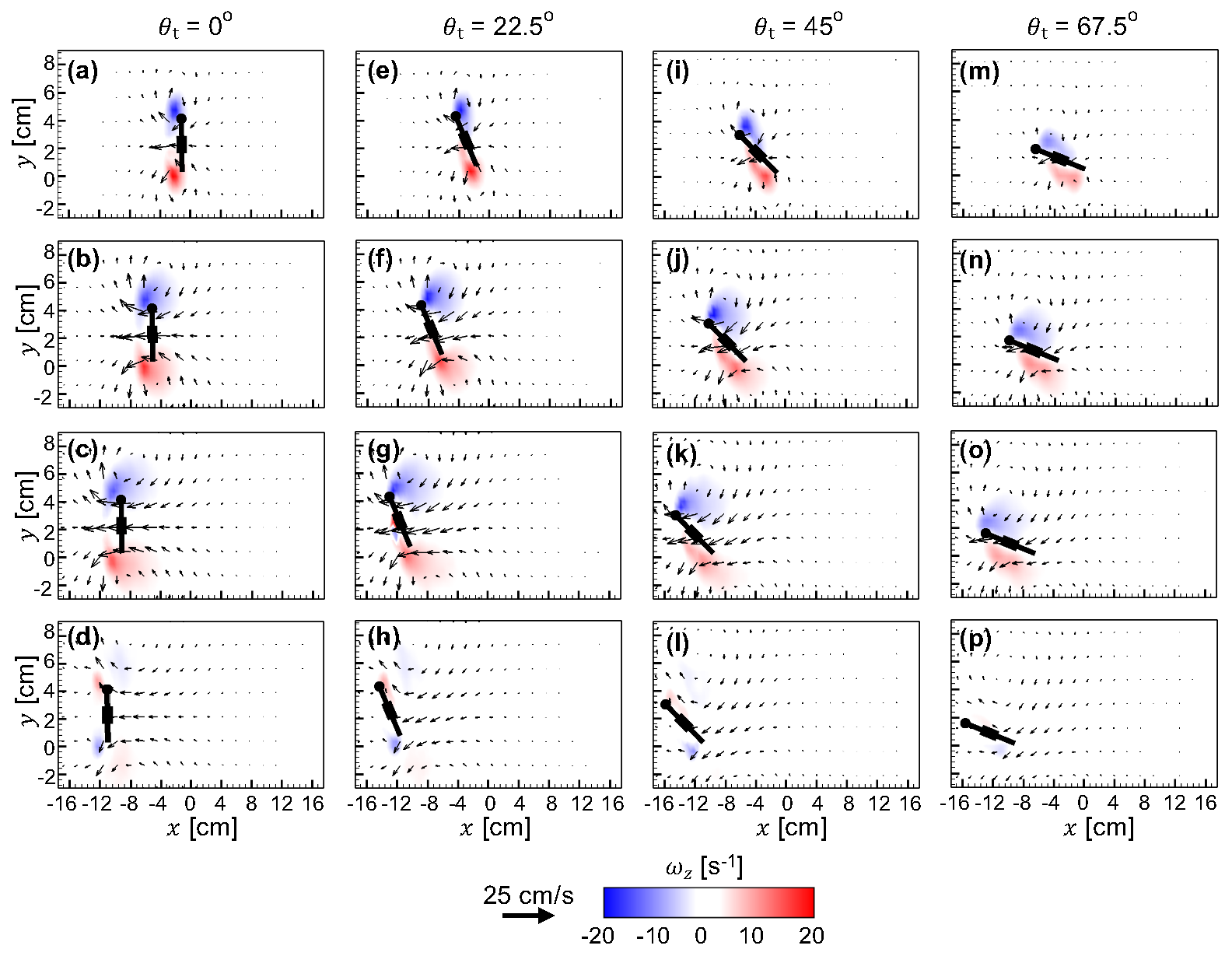}
    \caption{Velocity vectors overlaid on out-of-plane $z$-vorticity ($\omega_z$) contours for a single bristled wing in linear translation at \rey=10.~(a)-(d)~$\theta_\textnormal{t}$=0$^\circ$;~(e)-(h)~$\theta_\textnormal{t}$=22.5$^\circ$;~(i)-(l)~$\theta_\textnormal{t}$=45$^\circ$;~(m)-(p)~$\theta_\textnormal{t}$=67.5$^\circ$. For each $\theta_\textnormal{t}$, 4 timepoints (25\%, 50\%, 75\% and 100\% of cycle time) are shown along each column ((a)-(d); (e)-(h); (i)-(l); (m)-(p)) from top to bottom.
}
\label{fig10}
\end{figure*}

\noindent\underline{\bf\em Vorticity distribution.} A single bristled wing in linear translation produced counter-rotating vortices at the LE and TE ({\bf Figure~\ref{fig10}}). Across all $\theta_\text{t}$ values, we observed a LEV and a TEV that were attached to the wing, and their strength increased in time before dissipating at the end of the cycle (100\%$T$). Also, increasing $\theta_\text{t}$ decreased the strength of both the LEV and TEV during early translation ({\bf Figure~\ref{fig10}(a),(e),(i),(m)}). Minimal variation was observed in the vorticity magnitudes of LEV and TEV cores from 50\%$T$ to 75\%$T$ across all $\theta_\text{t}$ values.

\begin{figure*}[b]
    \centering
    \includegraphics[width=0.88\textwidth]{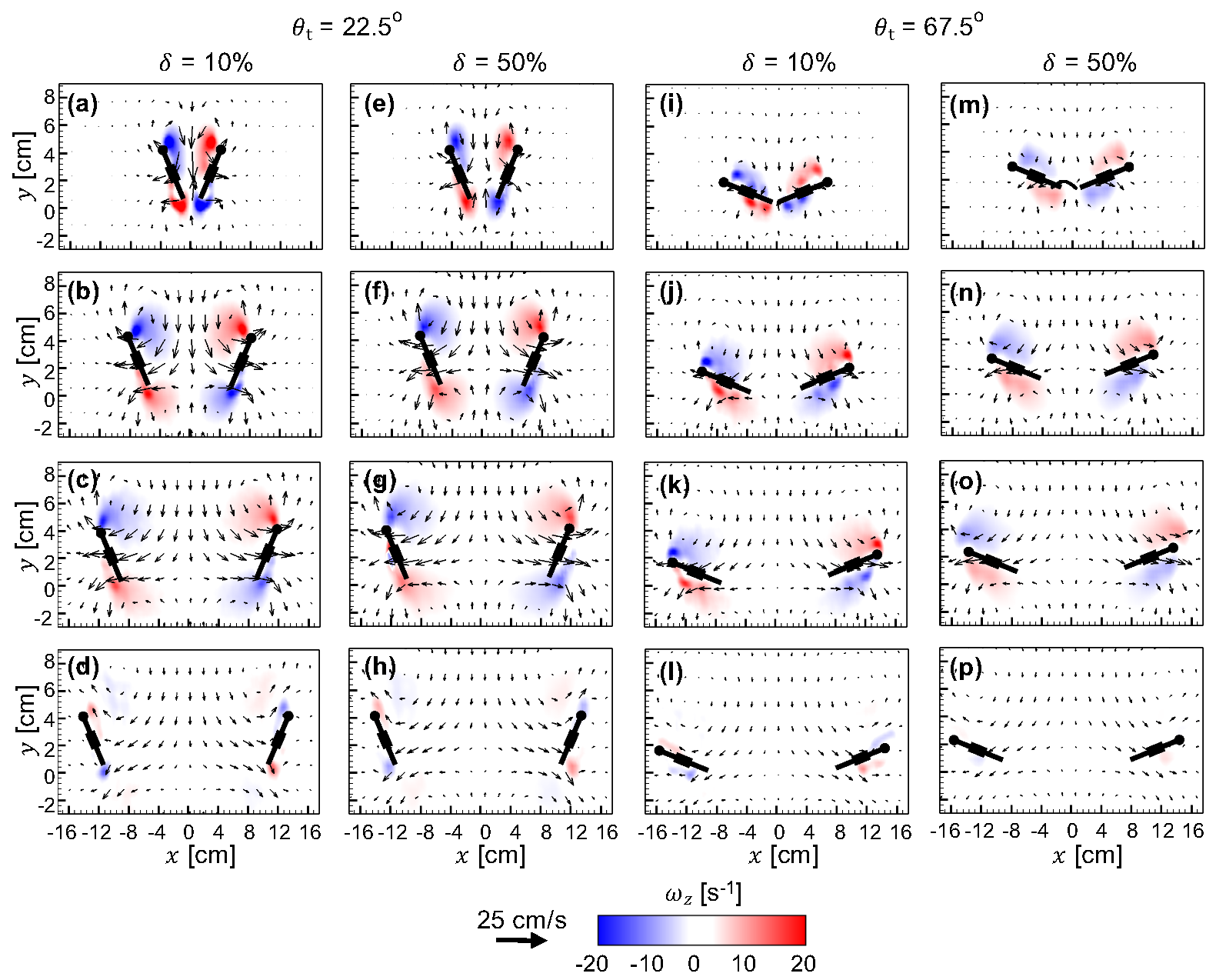}
    \caption{Velocity vectors overlaid on out-of-plane $z$-vorticity ($\omega_z$) contours for a bristled wing pair in linear translation at \rey=10.~$\theta_\textnormal{t}$=22.5$^\circ$ is shown for~$\delta$=10\% in (a)-(d) and for~$\delta$=50\% in (e)-(h).~$\theta_\textnormal{t}$=67.5$^\circ$ is shown for~$\delta$=10\% in (i)-(l) and for~$\delta$=50\% in (m)-(p).~For each $\theta_\textnormal{t}$, 4 timepoints (25\%, 50\%, 75\% and 100\% of cycle time) are shown along each column ((a)-(d); (e)-(h); (i)-(l); (m)-(p)) from top to bottom.
}
\label{fig11}
\end{figure*}

For a bristled wing pair in linear translation at $\theta_\text{t}$=22.5$^\circ$, increasing $\delta$ from 10\% to 50\% decreased the strength of both the LEV and TEV (compare {\bf Figure~\ref{fig11}(a)-(d)} and {\bf Figure~\ref{fig11}(e)-(h)}). However, at the end of cycle, vorticity distribution around each wing of the bristled wing pair was similar to that of a single wing in linear translation (compare {\bf Figure~\ref{fig10}(h)} and {\bf Figure~\ref{fig11}(d),(h)}). Similar to the single bristled wing in linear translation, we observed minimal variation in the vorticity magnitudes of the LEV and TEV from 50\%$T$ to 75\%$T$ ({\bf Figure~\ref{fig11}(b),(c),(f),(g)}). Similarly, for the bristled wing pair in linear translation at $\theta_\text{t}$=67.5$^\circ$, increasing $\delta$ decreased the strength of both the LEV and TEV (compare {\bf Figure~\ref{fig11}(i)-(l)} and {\bf Figure~\ref{fig11}(m)-(p)}). In contrast to $\theta_\text{t}$ = 22.5$^\circ$, LEV and TEV strength for $\theta_\text{t}$=67.5$^\circ$ showed larger variation with increasing $\delta$ throughout the cycle.

\begin{figure*}[b]
    \centering
    \includegraphics[width=0.88\textwidth]{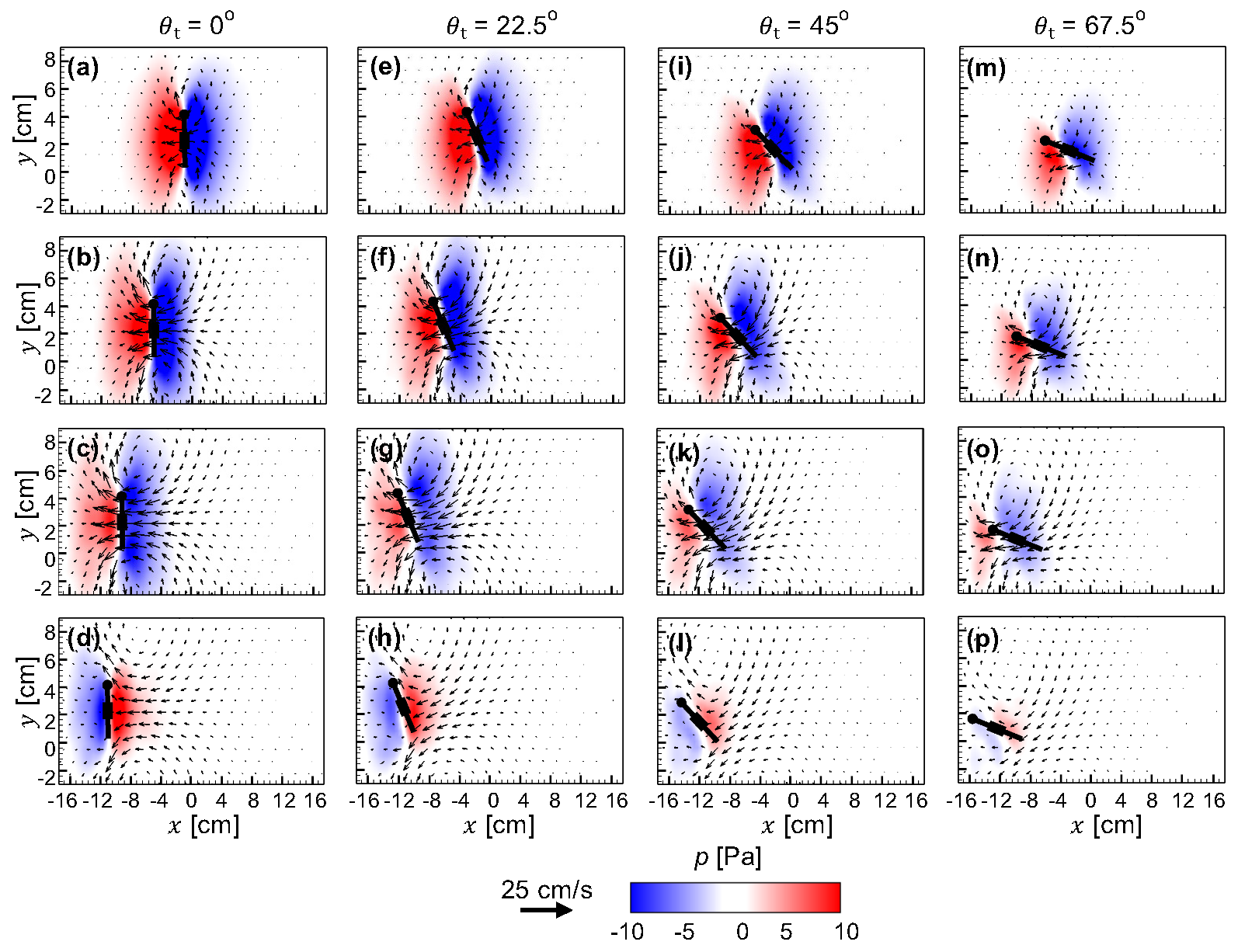}
    \caption{Velocity vectors overlaid on pressure ($p$) contours for a single bristled wing in linear translation at \rey=10.~(a)-(d)~$\theta_\textnormal{r}$=0$^\circ$;~(e)-(h)~$\theta_\textnormal{r}$=22.5$^\circ$;~(i)-(l)~$\theta_\textnormal{r}$=45$^\circ$;~(m)-(p)~$\theta_\textnormal{t}$=67.5$^\circ$.~For each $\theta_\textnormal{t}$, 4 timepoints (25\%, 50\%, 75\% and 100\% of cycle time) are shown along each column ((a)-(d); (e)-(h); (i)-(l); (m)-(p)) from top to bottom.
}
\label{fig12}
\end{figure*}

\noindent\underline{\bf\em Pressure distribution.} Similar to a single rotating wing, a single bristled wing undergoing linear translation showed positive and negative pressure regions below and above the wing, respectively ({\bf Figure~\ref{fig12}}). Time-variation of pressure distribution around the single translating wing was similar for all $\theta_\text{t}$ conditions.Increasing $\theta_\text{t}$ weakened the pressure distribution throughout the cycle. In addition, pressure distribution around the wing flipped in sign at the end of the translation (100\%$T$). This pressure reversal was more pronounced for smaller $\theta_\text{t}$ ($\leq$ 22.5$^\circ$).

\begin{figure*}[b]
    \centering
    \includegraphics[width=0.88\textwidth]{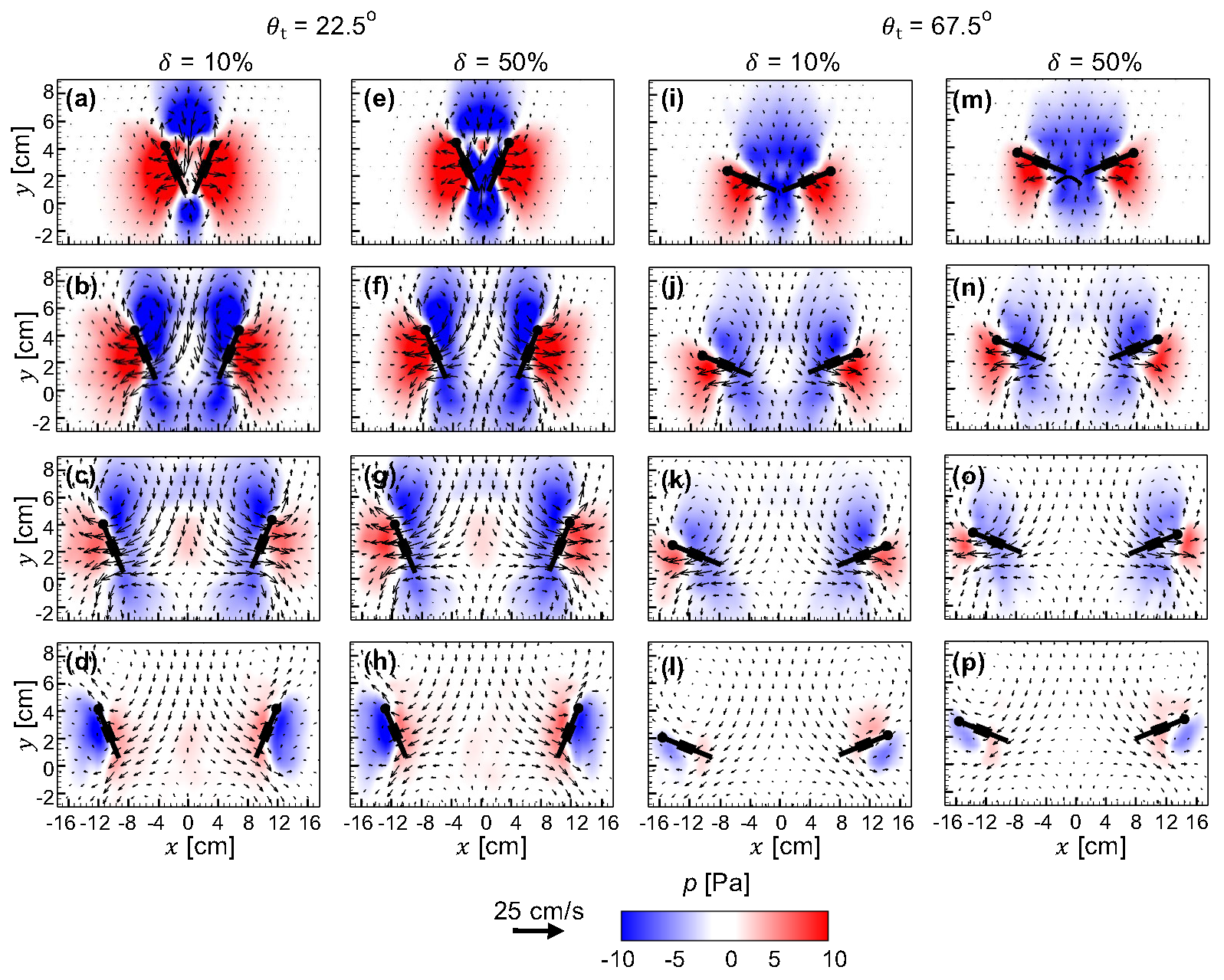}
    \caption{Velocity vectors overlaid on pressure ($p$) contours for a bristled wing pair in linear translation at \rey=10.~$\theta_\textnormal{t}$=22.5$^\circ$ is shown for~$\delta$=10\% in (a)-(d) and for~$\delta$=50\% in (e)-(h).~$\theta_\textnormal{t}$=67.5$^\circ$ is shown for~$\delta$=10\% in (i)-(l) and for~$\delta$=50\% in (m)-(p).~For each $\theta_\textnormal{t}$, 4 timepoints (25\%, 50\%, 75\% and 100\% of cycle time) are shown along each column ((a)-(d); (e)-(h); (i)-(l); (m)-(p)) from top to bottom.
}
\label{fig13}
\end{figure*}

Pressure distribution around a bristle wing pair in linear translation ({\bf Figure~\ref{fig13}}) was found to be different compared to that of a translating single wing ({\bf Figure~\ref{fig12}}) mostly at the start of the cycle on account of wing-wing interaction. During initial stages of linear translation, a diffused negative pressure region was observed near the LEs just above the cavity between the wings and near the TEs ({\bf Figure~\ref{fig13}(a),(e),(i),(m)}). Also, a diffused region of positive pressure was observed below each wing. For $\delta$ = 10$\%$ and $\theta_\text{t}$=22.5$^\circ$,  we observed a diffused region of negative pressure to be distributed in the cavity between the wing pair and near the LE at 50\%$T$ ({\bf Figure~\ref{fig13}(b)}). This is in contrast to the positive pressure region that was observed between the wing pair at the same time point during rotation to $\theta_\text{r}$=22.5$^\circ$ ({\bf Figure~\ref{fig8}(b)}). The magnitude of negative pressure in the cavity decreased with increasing time. Similar to the single translating wing, we observed the positive and negative pressure regions to flip positions at the end of the cycle (100\%$T$; {\bf Figure~\ref{fig13}(d),(h),(l),(p)}). Increasing $\delta$ to 50\% for $\theta_\text{t}$=22.5$^\circ$ reduced the negative pressure between the wings (compare {\bf Figure~\ref{fig13}(b)} and {\bf Figure~\ref{fig13}(f)}). From $\sim$50\%$T$ onward for $\theta_\text{t}$=22.5$^\circ$, we found both the positive and negative pressure distribution around the wing to be mostly unaffected with increasing $\delta$. 

In contrast to $\theta_\text{t}$=22.5$^\circ$, linear translation of the bristled wing pair at $\theta_\text{t}$=67.5$^\circ$ showed minimal change in pressure distribution when comparing identical time points at $\delta$=10\% ({\bf Figure~\ref{fig13}(i)-(l)}) and $\delta$=50\% ({\bf Figure~\ref{fig13}(m)-(p)}). This suggests that there is a limit to $\theta_\text{t}$ after which wing-wing interaction is unaltered for $\delta\geq$10\%.  Just after the start of translation at $\theta_\text{t}$=67.5$^\circ$, we found negative pressure to be distributed in between the wing and positive pressure below the wings for both $\delta$=10\% and 50\%. 
The magnitudes of negative and positive pressures at $\theta_\text{t}$=67.5$^\circ$ were found to be substantially lower than those of $\theta_\text{t}$=22.5$^\circ$ throughout the cycle.

\subsection{Bristled wings during combined rotation and linear translation}
\begin{figure*}[b]
    \centering
    \includegraphics[width=0.8\textwidth]{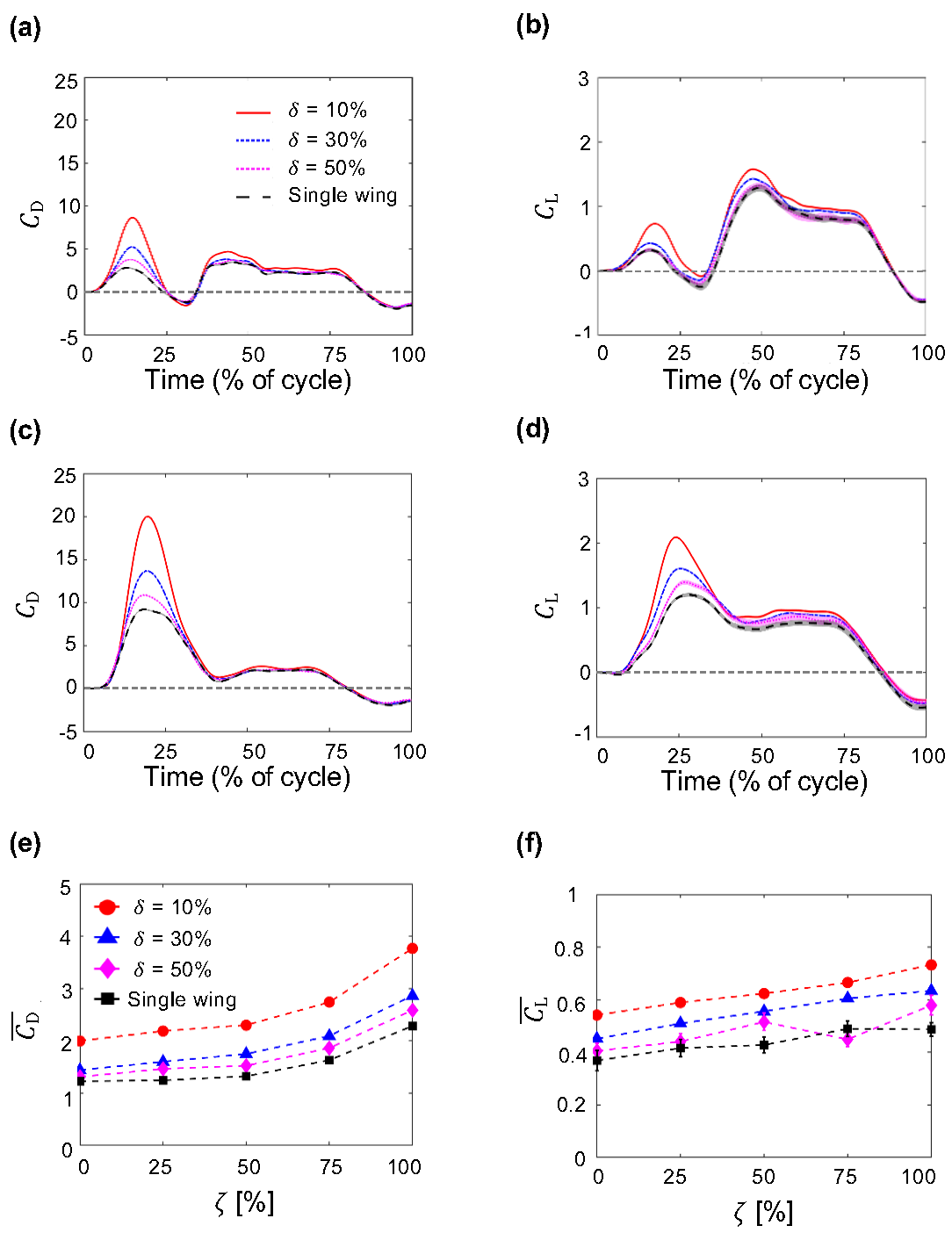}
    \caption{Force coefficients during combined rotation and linear translation of bristled wings at \rey=10. Shading around each curve represents $\pm$1 SD across 30 cycles.~(a) and (b) show time-variation of $C_\textnormal{D}$ and $C_\textnormal{L}$, respectively, for overlap $\zeta$=25\%.~(c) and (d) show time-variation of $C_\textnormal{D}$ and $C_\textnormal{L}$, respectively, for $\zeta$=50\%.~(e) and (f) show cycle-averaged coefficients $\overline{C_\textnormal{D}}$ and $\overline{C_\textnormal{L}}$, respectively, for varying $\zeta$.~Legend for (b)-(d) is shown in (a); legend for (f) is shown in (e).
}
\label{fig14}
\end{figure*}

\noindent\underline{\bf\em Aerodynamic force generation.} At $\zeta$=25\%, both $C_\text{L}$ and $C_\text{D}$ were found to peak at two timepoints in the cycle ({\bf Figure~\ref{fig14}(a),(b)}). One of the timepoints correspond to where the rotational wing motion reached peak velocity and other time point correspond to the peak translational velocity. With increase in $\zeta$ to 100\% ({\bf Figure~\ref{fig14} (c),(d)}), we observed both $C_\text{L}$ and $C_\text{D}$ to peak at only one time point early in the cycle. In addition, peak values of $C_\text{L}$ and $C_\text{D}$ increased with increasing $\zeta$. For each $\zeta$, increasing $\delta$ decreased peak values of both $C_\text{L}$ and $C_\text{D}$. However, during wing translation following overlapping motion, both $C_\text{L}$ and $C_\text{D}$ showed minimal variation for varying $\delta$. Similar to linear translation, both $C_\text{L}$ and $C_\text{D}$ dropped below zero close towards the end of the cycle.

In general, cycle-averaged coefficients ($\overline{C_\text{D}}$ and $\overline{C_\text{L}}$, {\bf Figure~\ref{fig14}(e),(f)}) were observed to increase with increasing $\zeta$. Increasing $\delta$ decreased both $\overline{C_\text{D}}$ and $\overline{C_\text{L}}$. The extent of $\overline{C_\text{L}}$ variation with $\zeta$ was substantially smaller than that of $\overline{C_\text{D}}$.

\begin{figure*}[h]
    \centering
    \includegraphics[width=0.88\textwidth]{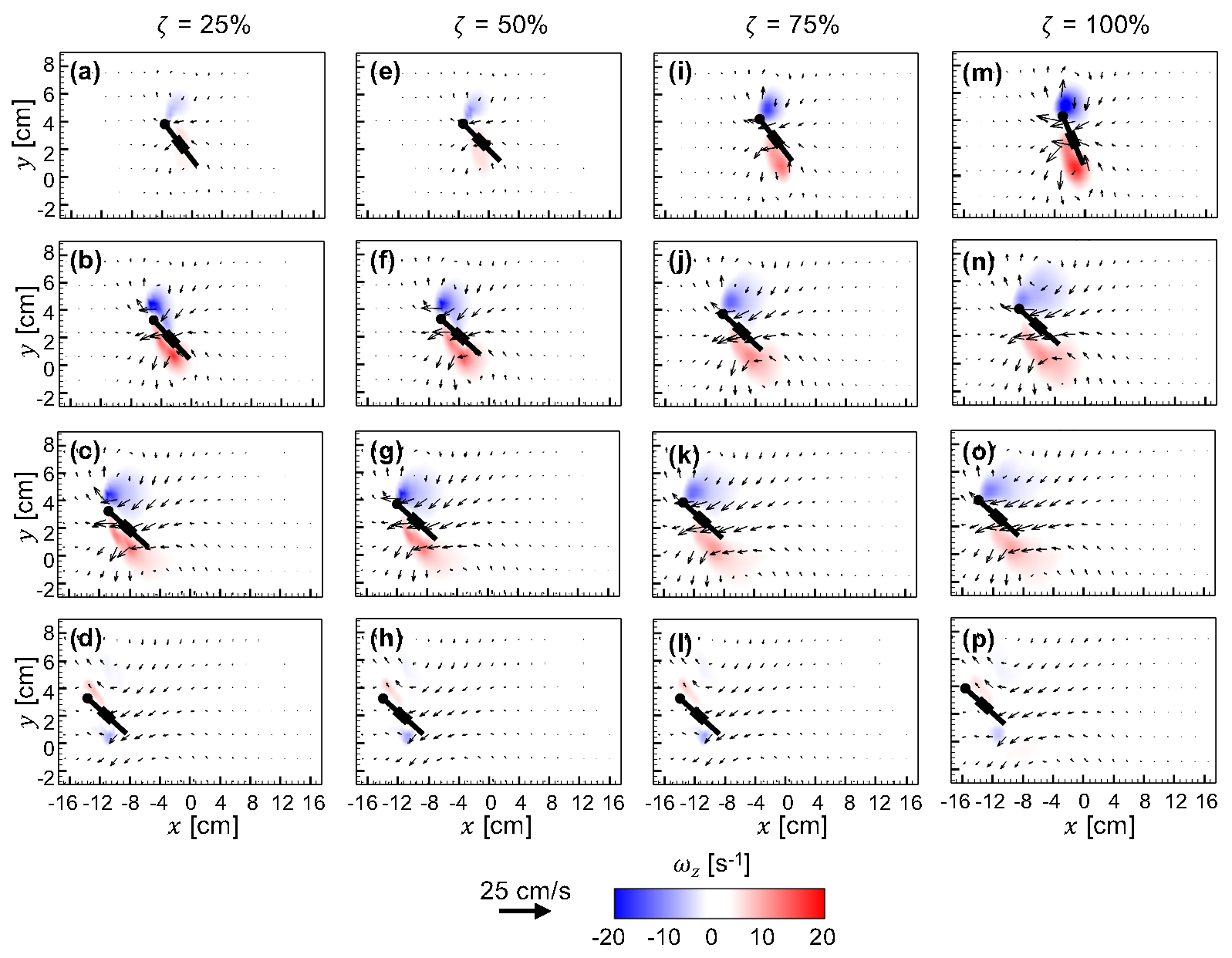}
    \caption{Velocity vectors overlaid on out-of-plane $z$-vorticity ($\omega_z$) contours for combined rotation and linear translation of a single bristled wing at \rey=10.~(a)-(d)~$\zeta$=25\%;~(e)-(h)~$\zeta$=50\%;~(i)-(l)~$\zeta$=75\%;~(m)-(p)~$\zeta$=100\%. For each $\zeta$, 4 timepoints (25\%, 50\%, 75\% and 100\% of cycle time) are shown along each column ((a)-(d); (e)-(h); (i)-(l); (m)-(p)) from top to bottom.
}
\label{fig15}
\end{figure*}

\noindent\underline{\bf\em Vorticity distribution.} {\bf Figure~\ref{fig15}} shows the flow generated by a single bristled wing performing combined rotation and linear translation. With increasing $\zeta$, the strength of both LEV and TEV were found to increase during early stages of wing motion (25\%$T$) 
This could likely be on account of both wings reaching rotational deceleration phase at 25\%$T$ for all $\zeta$. At 75\%$T$, the strength of both LEV and TEV were found to have little to no change with increasing $\zeta$ ({\bf Figure~\ref{fig14}(c),(g),(k),(o)}).

\begin{figure*}[b]
    \centering
    \includegraphics[width=0.88\textwidth]{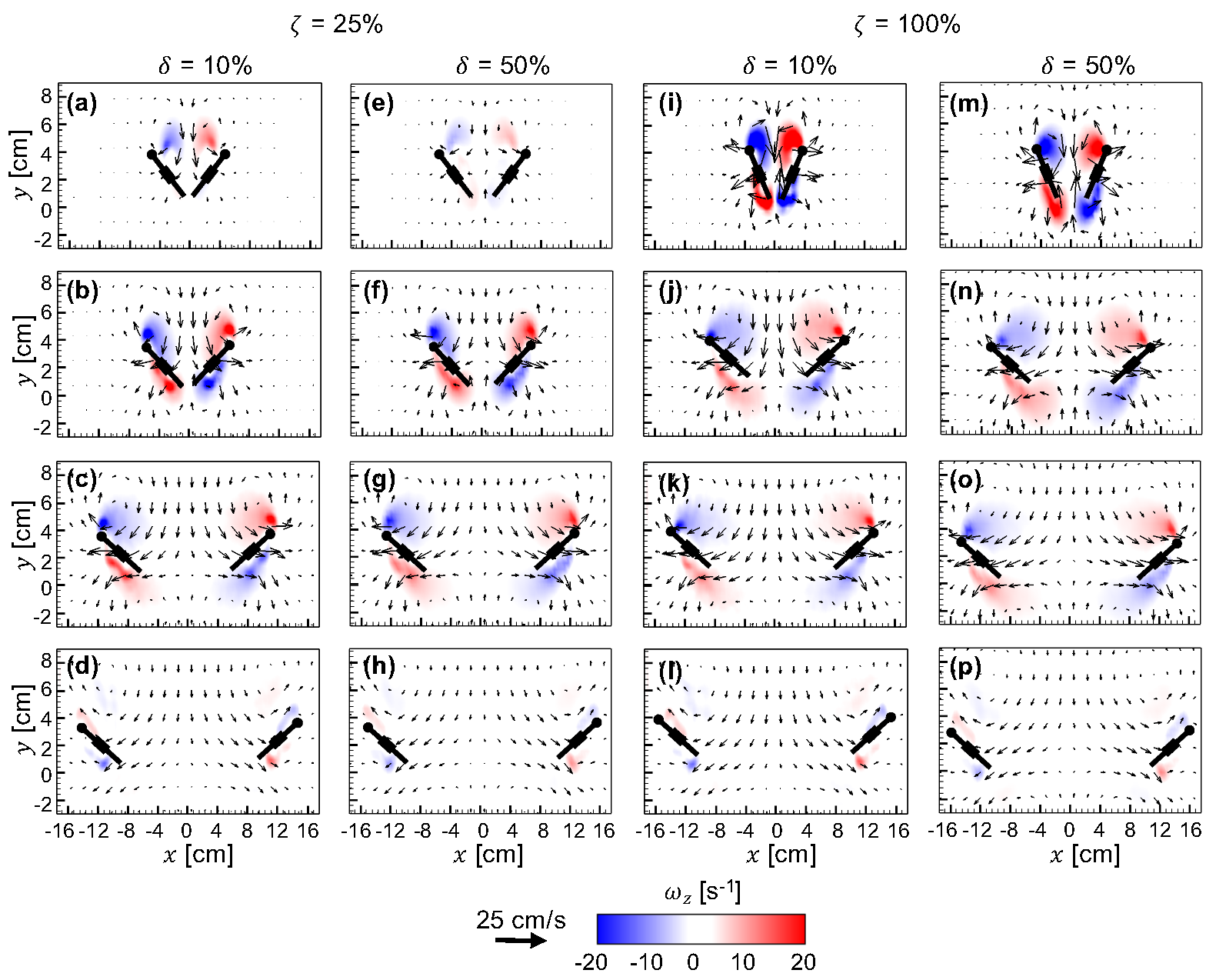}
    \caption{Velocity vectors overlaid on out-of-plane $z$-vorticity ($\omega_z$) contours for combined rotation and linear translation of a bristled wing pair at \rey=10.~$\zeta$=25\% is shown for~$\delta$=10\% in (a)-(d) and for~$\delta$=50\% in (e)-(h).~$\zeta$=100\% is shown for~$\delta$=10\% in (i)-(l) and for~$\delta$=50\% in (m)-(p).~For each $\zeta$, 4 timepoints (25\%, 50\%, 75\% and 100\% of cycle time) are shown along each column ((a)-(d); (e)-(h); (i)-(l); (m)-(p)) from top to bottom.
}
\label{fig16}
\end{figure*}

For a bristled wing pair performing combined rotation and linear translation at $\zeta$=25\% ({\bf Figure~\ref{fig16}(a)-(h)}), increasing $\delta$ decreased the strength of both the LEV and TEV during initial stages of wing motion (25\%$T$ and 50\%$T$). Towards the end of cycle with increasing $\delta$, there were essentially no changes to the vorticity of the LEV and TEV cores. Similar trends were also observed for $\zeta$=100\% ({\bf Figure~\ref{fig16}(i)-(p)}). 

Similar to a single wing, increasing the overlap ($\zeta$) for one particular initial inter-wing spacing ($\delta$) increased the strength of both LEV and TEV at 25$\%$ and 50$\%$ of cycle time. However, LEV and TEV strength showed little to no variations towards the end of cycle time for $\zeta$ = 25$\%$ and 100$\%$.

\begin{figure*}[b]
    \centering
    \includegraphics[width=0.88\textwidth]{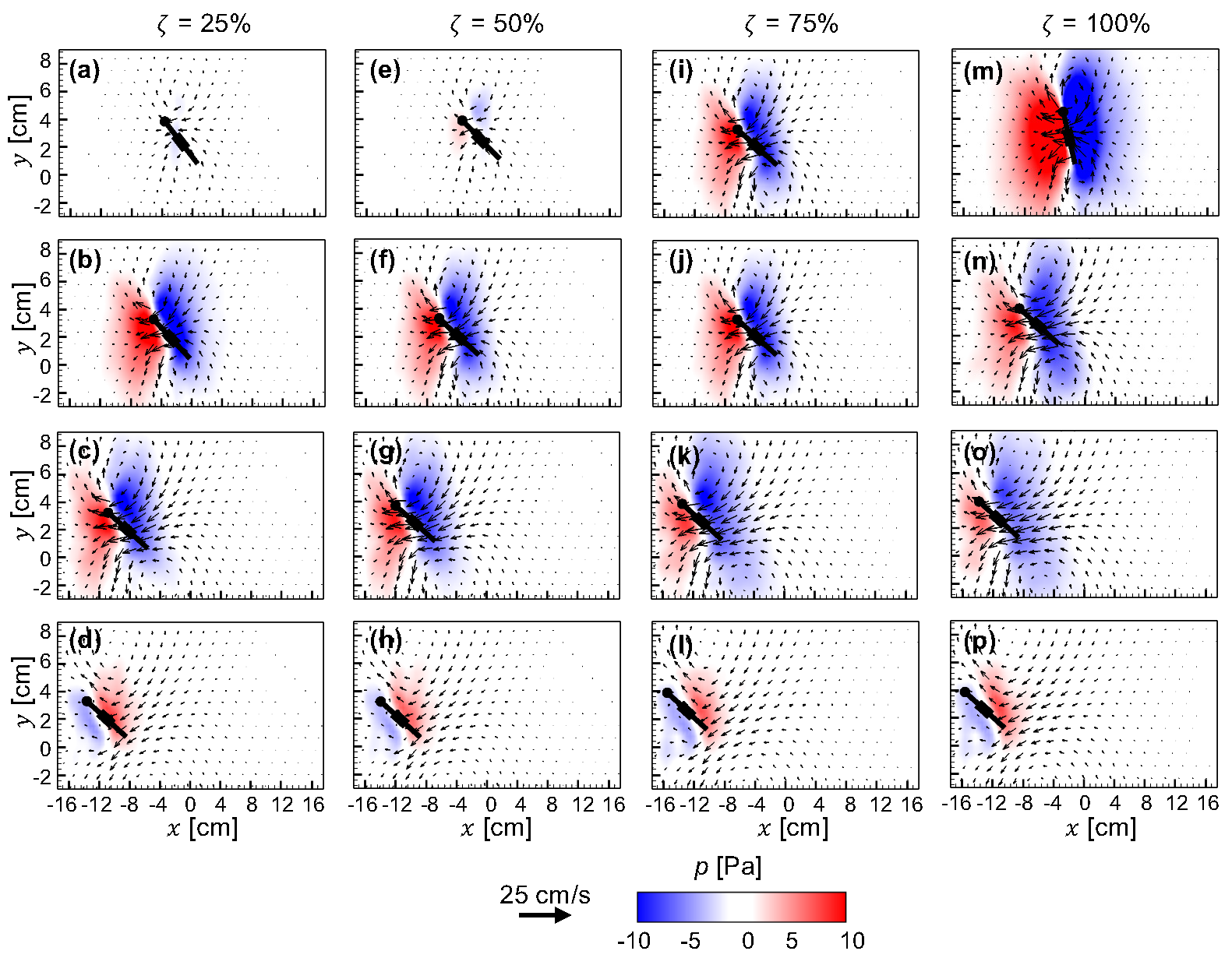}
    \caption{Velocity vectors overlaid on pressure ($p$) contours for combined rotation and linear translation of a single bristled wing at \rey=10.~(a)-(d)~$\zeta$=25\%;~(e)-(h)~$\zeta$=50\%;~(i)-(l)~$\zeta$=75\%;~(m)-(p)~$\zeta$=100\%.~For each $\zeta$, 4 timepoints (25\%, 50\%, 75\% and 100\% of cycle time) are shown along each column ((a)-(d); (e)-(h); (i)-(l); (m)-(p)) from top to bottom.}
\label{fig17}
\end{figure*}

\noindent\underline{\bf\em Pressure distribution.} A single bristled wing performing combined rotation and linear translation showed substantial changes in pressure distribution with changing $\zeta$ (Figure~\ref{fig17}). Similar to vorticity distribution, both positive and negative pressure magnitudes increased with increasing overlap during 25\%$T$ ({\bf Figure~\ref{fig17}(a),(e),(i),(m)}) and 50\%$T$ ({\bf Figure~\ref{fig17}(b),(f),(j),(n)}). At 75\%$T$ ({\bf Figure~\ref{fig17}(c),(g),(k),(o)}) and 100\%$T$ ({\bf Figure~\ref{fig17}(d),(h),(l),(p)}), increasing $\zeta$ resulted in little to no changes to the pressure distribution around the wing.

\begin{figure*}[b]
    \centering
    \includegraphics[width=0.88\textwidth]{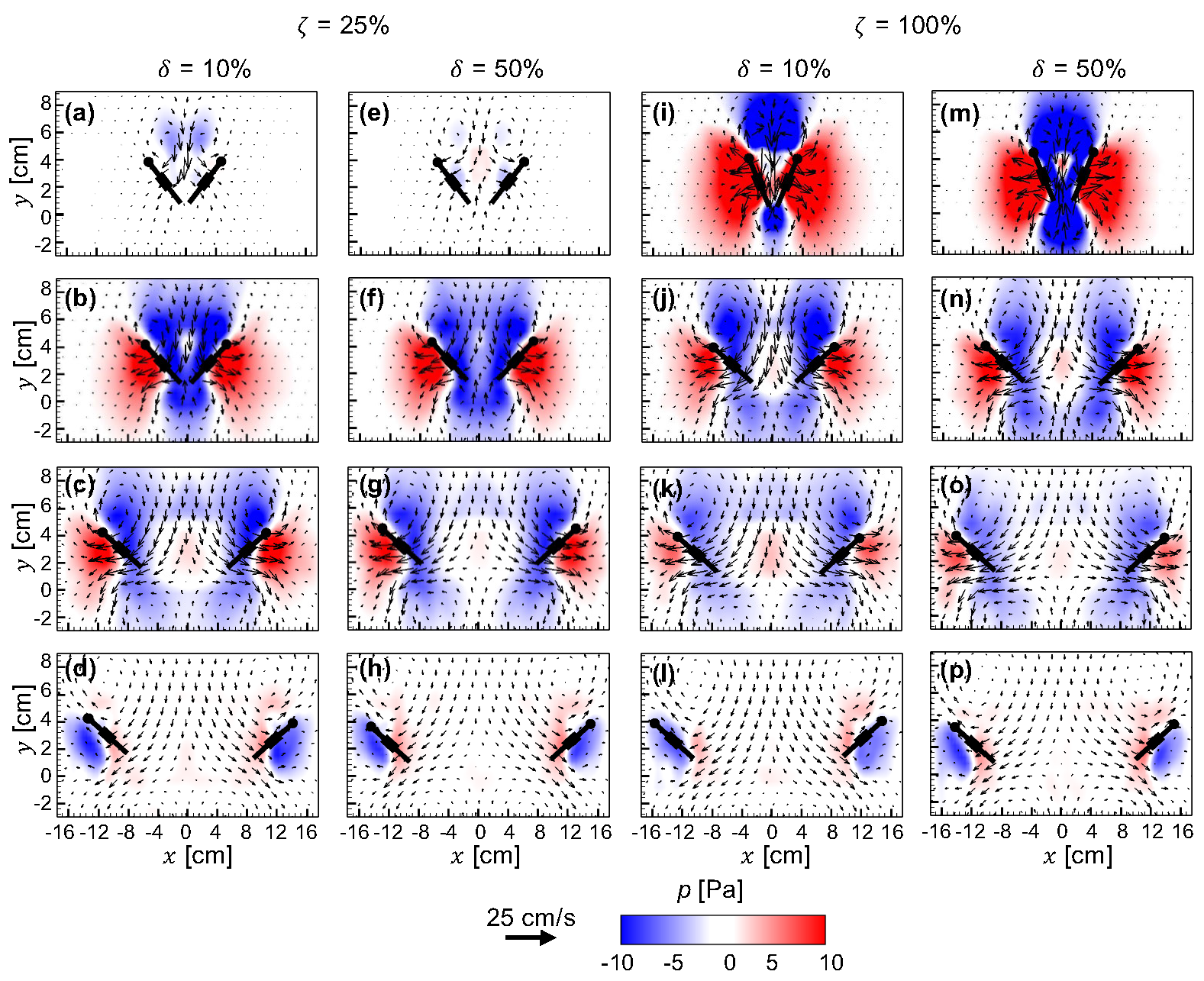}
    \caption{Velocity vectors overlaid on pressure ($p$) contours for combined rotation and linear translation of a bristled wing pair at \rey=10.~$\zeta$=25\% is shown for~$\delta$=10\% in (a)-(d) and for~$\delta$=50\% in (e)-(h).~$\zeta$=100\% is shown for~$\delta$=10\% in (i)-(l) and for~$\delta$=50\% in (m)-(p).~For each $\theta_\textnormal{t}$, 4 timepoints (25\%, 50\%, 75\% and 100\% of cycle time) are shown along each column ((a)-(d); (e)-(h); (i)-(l); (m)-(p)) from top to bottom.
}
\label{fig18}
\end{figure*}

Pressure distribution around a bristle wing pair ({\bf Figure~\ref{fig18}}) was found to be different compared to that of a single wing (both cases performing rotation and linear translation) mostly during early stages of the cycle, where wing-wing interaction appears to have the most influence. During the earlier part of  the combined rotation and translation cycle at 50\%$T$ and $\zeta$=25\% ({\bf Figure~\ref{fig18}(b),(f)}), we observed an increase in negative pressure distribution within the cavity between the wings and positive pressure distributed below each wing. With further increase in time from 75\%$T$ ({\bf Figure~\ref{fig18}(c),(g)}) to 100\%$T$ ({\bf Figure~\ref{fig18}(d),(h)}), the pressure distribution starts to closely resemble that of a single wing, suggesting diminished influence of wing-wing interaction. Increasing $\delta$ at $\zeta$ = 25$\%$ resulted in a drop in the pressure distribution only during the start of the cycle (25\%$T$; {\bf Figure~\ref{fig18}(a),(e)}), and minimal variation in pressure distribution was observed between $\delta$=10\% ({\bf Figure~\ref{fig18}(b)-(d)}) and $\delta$=50\% ({\bf Figure~\ref{fig18}(f)-(h)}) for the remainder of the cycle.

 Similar trends were observed with increasing $\delta$ for $\zeta$=100\% ({\bf Figure~\ref{fig18}(i)-(p)}) as compared to those discussed for $\zeta$=25\%. However, we observed the development of a strong negative pressure region in the cavity between the wings for $\delta$=50\% early into the cycle (25\%$T$; {\bf Figure~\ref{fig18}(m)}). Also, larger negative and positive regions were observed for $\zeta$=100\% as compared to $\zeta$=25\%. However, we did not observe noticeable differences in the pressure distribution at 75\%$T$ and 100\%$T$ when changing either $\zeta$ or $\delta$.  

\begin{figure*}[b]
    \centering
    \includegraphics[width=0.8\textwidth]{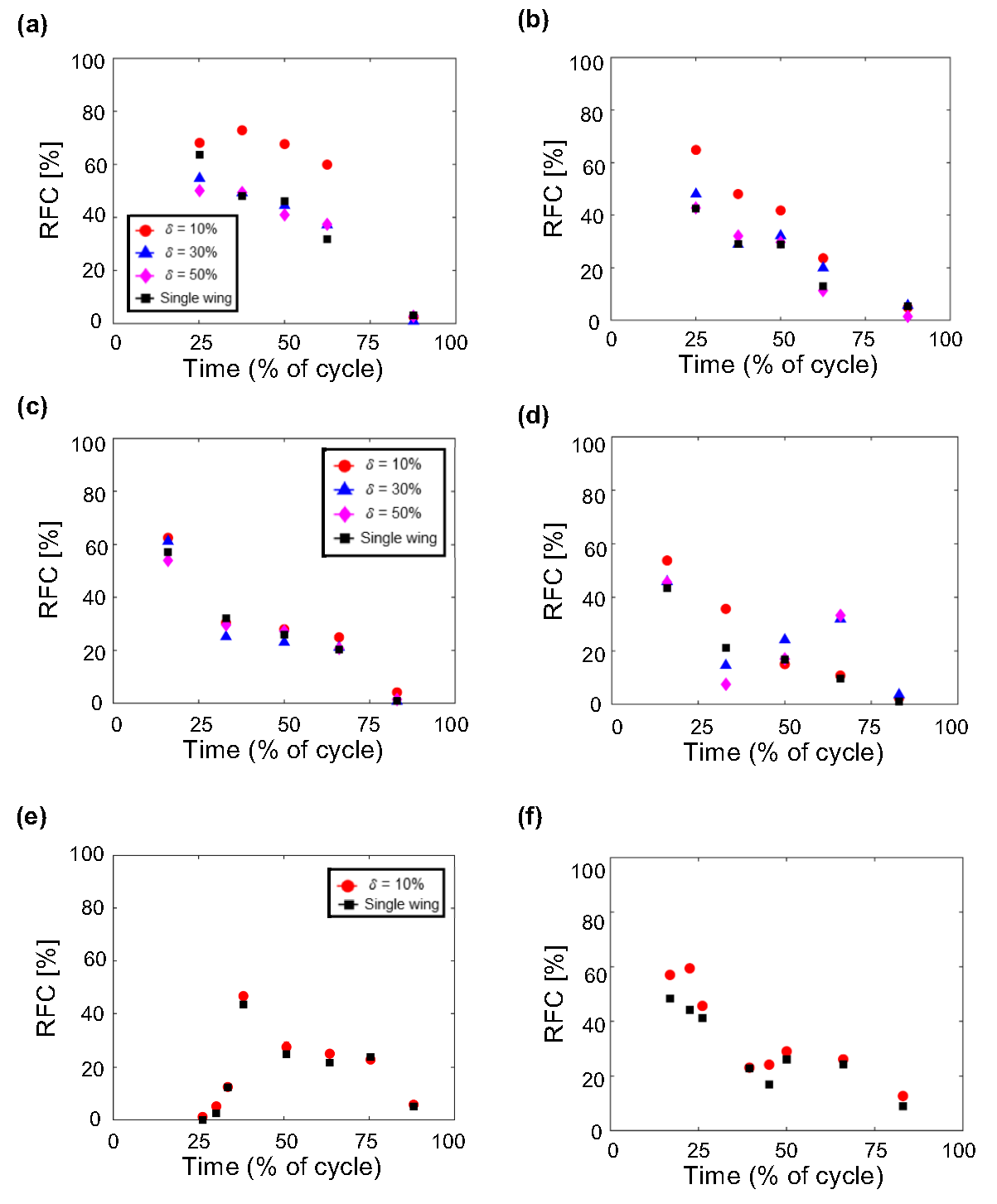}
    \caption{Time-variation of reverse flow capacity (RFC), characterizing the reduction in volumetric flow of a bristled wing (or wing pair) with respect to a geometrically equivalent solid wing, as a function of $\delta$ and wing kinematics.~(a) and (b) show RFC during rotation at $\theta_\textnormal{r}$=22.5$^\circ$ and $\theta_\textnormal{r}$=67.5$^\circ$, respectively.~(c) and (d) show RFC during linear translation at $\theta_\textnormal{t}$=22.5$^\circ$ and $\theta_\textnormal{t}$=67.5$^\circ$, respectively.~(e) and (f) show RFC during combined rotation and linear translation at $\zeta$=25\% and $\zeta$=100\%, respectively. Both single bristled wing and bristled wing pairs are included. See subsection~\ref{sec:Calculated quantities} for more details on definition and calculation of RFC.}
\label{fig22}
\end{figure*}

\subsection{Reverse flow through bristled wings}
Reverse flow capacity (RFC) by a bristled wing was quantified using the equation~\ref{eq:FlowReduction}. RFC gives a dimensionless estimate of the capability of a given bristled wing model to leak fluid through the bristles on a bristled wing model for varying $\delta$, $\theta_\text{t}$, $\theta_\text{r}$, and $\zeta$ (Figure~\ref{fig22}). For all $\theta_\text{r}$, RFC was in the range of 0\%-80\% ({\bf Figure~\ref{fig22}(a),(b)}). RFC was larger for smaller $\theta_\text{r}$ of 22.5$^\circ$ as compared to 67.5$^\circ$ at the same \% of cycle time. In addition, having the wings closer ($\delta$=10$\%$) showed higher RFC for $\theta_\text{r}$=22.5$^\circ$. This is in agreement with the results of Loudon \etal~\citep{Loudon94}, where the presence of a wall near bristled appendages was observed to promote inter-bristle flow. This increase in RFC can be attributed to net changes in pressure distribution around the wing for $\delta$ = 10\% at $\theta_\text{r}$=22.5$^\circ$. 

Increasing $\delta$ beyond 10\% showed little to no change in RFC. In addition, for changing $\theta_\text{t}$ ({\bf Figure~\ref{fig22}(c),(d)}) and $\zeta$ ({\bf Figure~\ref{fig22}(e),(f)}), we observe very little variation in RFC across all $\delta$ values. However, the RFC was found to change in time for each $\theta_\text{t}$ or $\zeta$ (in addition to $\theta_\text{r}$). The latter suggests that RFC is largely dependent on wing kinematics and found to be more for smaller $\delta$. Interestingly, higher values of RFC that were observed for lower $\theta_\text{r}$      and smaller $\delta$ were also associated with large $C_\text{D}$. While it is intuitive to expect that a bristled wing with larger capacity to leak flow through the bristles will reduce drag, this counter-intuitive finding suggests that the high drag forces were generated by formation of shear layers around the bristles as has been noted in previous studies~\cite{Lee17,Kasoju18}.

\section{Discussion}
While several computational studies~\citep{Miller05,Miller09,Arora14,MaoSun2003,MaoSun06} have examined wing-wing interaction in fling at low \rey~for varying $\delta$ and $\zeta$, the wings were modeled as solid wings unlike the bristled wings typically seen in tiny flying insects. Further, the few computational studies of wing-wing interaction of bristled wings~\cite{Santhanakrishnan14,Jones16} did not isolate the specific roles of wing rotation from translation. We experimentally examined the flow structures and forces generated by a single bristled wing and a bristled wing pair under varying initial inter-wing distance ($\delta$) at \rey=10, for the following kinematics: rotation to $\theta_\text{r}$ about the TE, linear translation at a fixed angle $\theta_\text{t}$, and combined rotation and linear translation (overlap duration $\zeta$ in \%). The central findings for varying wing kinematics are: (1) increasing $\theta_\text{r}$ decreased both cycle-averaged lift ($\overline{C_\text{L}}$) and drag ($\overline{C_\text{D}}$) coefficients; (2) increasing $\theta_\text{t}$ decreased $\overline{C_\text{D}}$ and approached $\overline{C_\text{D}}$ of a single wing at $\theta_\text{t}$=67.5$^\circ$; (3) $\overline{C_\text{L}}$ increased with increasing $\theta_\text{t}$, peaking at $\theta_\text{t}$=45$^\circ$ and decreasing thereafter at $\theta_\text{t}$=45$^\circ$; and (4) increasing $\zeta$ increased both $\overline{C_\text{L}}$ and $\overline{C_\text{D}}$. For all wing kinematics examined here, increasing $\delta$ resulted in a disproportionally smaller reduction of $\overline{C_\text{L}}$ as compared to larger reduction of $\overline{C_\text{D}}$. We find that peak ${C_\text{L}}$ of a wing pair separated by $\delta$=10\% during rotation and during combined rotation and linear translation ($\zeta$=25$\%$) occurs close to the time point where an attached, asymmetric (in size) LEV-TEV pair was observed over the wing. Finally, large values of $C_\text{D}$ during rotation of a wing pair with $\delta$=10\% resulted from large positive pressure distribution between the wings.
 
\subsection{Implications of vorticity distribution on lift force generation}
\begin{figure*}[b]
    \centering
    \includegraphics[width=0.8\textwidth]{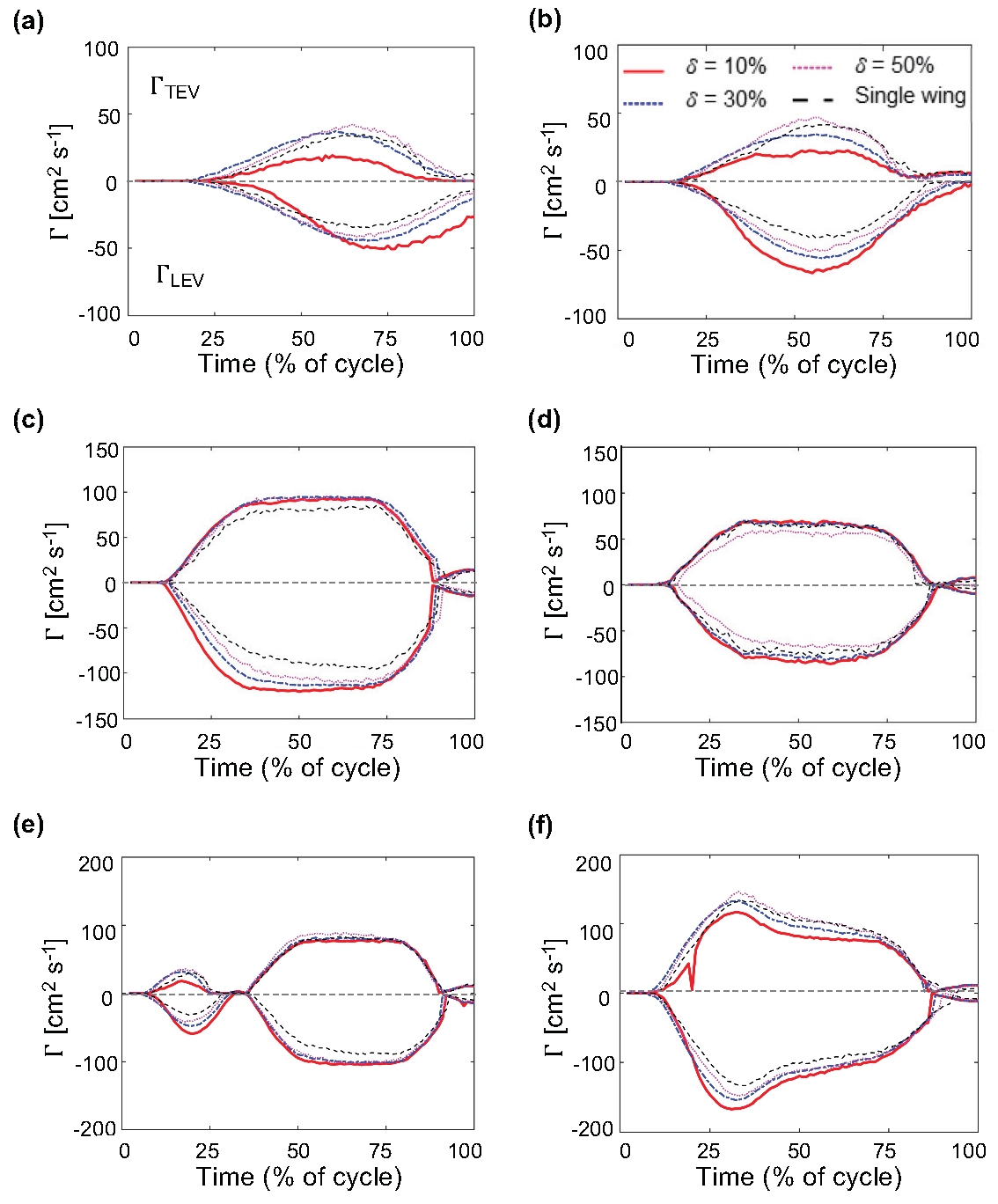}
    \caption{Circulation ($\Gamma$) of the leading edge vortex (LEV) and the trailing edge vortex (TEV) as a function of $\delta$ and wing kinematics.~(a) and (b) show $\Gamma$ during rotation at $\theta_\textnormal{r}$=22.5$^\circ$ and $\theta_\textnormal{r}$=67.5$^\circ$, respectively. (c) and (d) show $\Gamma$ during linear translation at $\theta_\textnormal{t}$=22.5$^\circ$ and $\theta_\textnormal{t}$=67.5$^\circ$, respectively. (e) and (f) show $\Gamma$ during combined rotation and linear translation at $\zeta$= 25\% and $\zeta$=100\%, respectively. Positive $\Gamma$  corresponds to TEV and negative $\Gamma$ corresponds to LEV. Both single bristled wing and bristled wing pairs are included. For bristled wing pairs, $\Gamma$ was only calculated on the left-wing. See subsection~\ref{sec:Calculated quantities} for more details on definition and calculation of $\Gamma$. 
}
\label{fig19}
\end{figure*}

Previous studies examining aerodynamic effects of varying $\delta$ of solid wing pairs~\citep{Arora14,MaoSun2003, MaoSun06} and porous wing pairs~\cite{Santhanakrishnan14} did not elaborate on the physical mechanism(s) responsible for lift augmentation observed with decreasing $\delta$. A stable, attached TEV has been observed in addition to the LEV for a single wing in revolution and in linear translation at \rey$\leq$32~\citep{Miller04,Santhanakrishnan18}, and this LEV-TEV `vortical symmetry' has been identified as a primary reason for diminished lift generation at this \rey~range~\cite{Miller04}. Miller and Peskin~\cite{Miller05} identified `vortical asymmetry' (larger LEV, smaller TEV) during fling of a solid wing pair at \rey$\leq$32 as the mechanism underlying the observed lift augmentation, suggesting that wing-wing interaction can help recover some of the lift lost during the remainder of the cycle (latter attributed to `vortical symmetry'). We examined circulation ($\Gamma$) of the LEV and TEV on a wing of the interacting bristled wing pair to explain the observed changes in lift generation under varying $\delta$ and kinematics ({\bf Figure~\ref{fig19}}).

\begin{figure*}[b]
	\centering
	\includegraphics[width=0.8\textwidth]{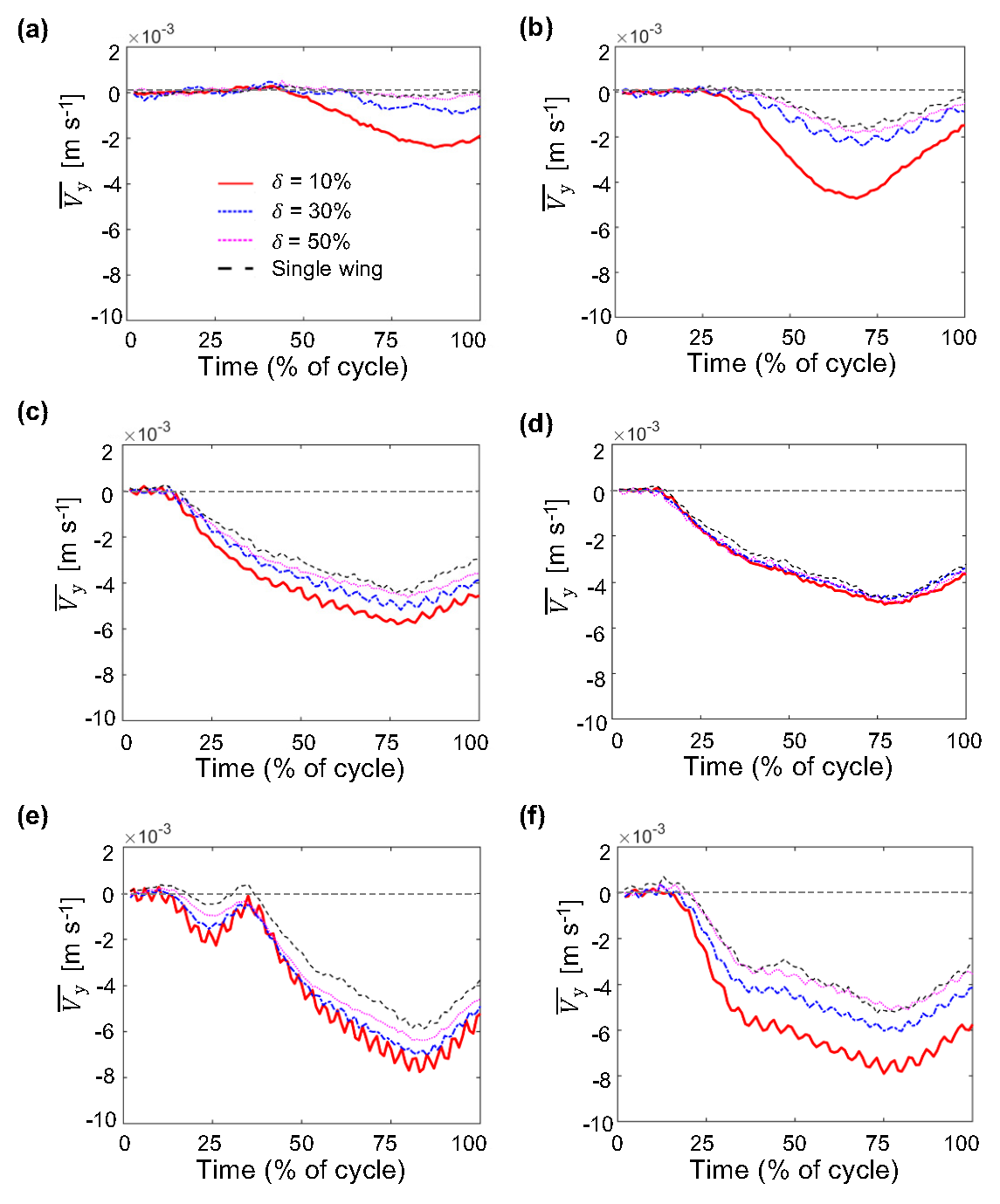}
	\caption{Time-variation of downwash ($\overline{V_y}$), defined as the average velocity of the flow displaced vertically downward due to wing motion, as a function of $\delta$ and wing kinematics.~(a) and (b) show $\overline{V_y}$ during rotation at $\theta_\textnormal{r}$=22.5$^\circ$ and $\theta_\textnormal{r}$=67.5$^\circ$, respectively. (c) and (d) show $\overline{V_y}$ during linear translation at $\theta_\textnormal{t}$=22.5$^\circ$ and $\theta_\textnormal{t}$=67.5$^\circ$, respectively. (e) and (f) show $\overline{V_y}$ during combined rotation and linear translation at $\zeta$= 25\% and $\zeta$=100\%, respectively. Both single bristled wing and bristled wing pairs are included. See subsection~\ref{sec:Calculated quantities} for more details on definition and calculation of downwash. 
}
	\label{fig20}
\end{figure*}

Increasing $\theta_\text{r}$ from 22.5$^\circ$ to 67.5$^\circ$ increased the peak net circulation on the wing ($|\Gamma_\text{LEV}|$-$|\Gamma_\text{TEV}|$) by roughly 2.5 times for $\delta$=10$\%$ ({\bf Figure~\ref{fig19}(a),(b)}). Surprisingly, we saw a drop in peak $C_\text{L}$ with increasing $\theta_\text{r}$ ({\bf Figure~\ref{fig4}(b),(d)}). To examine the reason for this discrepancy, we calculated the average downwash velocity ($\overline{V_{y}}$) ({\bf Figure~\ref{fig20}}). We observed a substantial increase in $\overline{V_{y}}$ with increased $\theta_\text{r}$. An increase in downwash velocity lowers the effective angle of attack~\cite{Sane03}, which can explain the observed reduction in peak $C_\text{L}$ with increasing $\theta_\text{r}$. Also, increasing $\theta_\text{r}$ shifted the formation of peak net circulation to occur early in time, similar to what we observed for peak $C_\text{L}$ with increasing $\theta_\text{r}$ ({\bf Figure~\ref{fig4}(d)}) . This was likely on account of the longer time scale for $\theta_\text{r}$=67.5$^\circ$ (compared to 22.5$^\circ$), enabling the LEV and TEV to develop in time. These results suggest that rotational motion continuously change the circulation around the wing by diffusing the LEV and TEV to remain attached in time. Increasing $\delta$ above 10\% resulted in lower variation of $C_\text{L}$ as well as net circulation around the wing. We see wing-wing interaction effects to diminish for $\delta$$>$50\%, thereby behaving like a single wing, which is in agreement with previous studies~\citep{MaoSun06,Arora14}. 

Increasing the translation angle ($\theta_\text{t}$) from 22.5$^\circ$ to 67.5$^\circ$ for $\delta$=10\% decreased the net circulation by 37\% ({\bf Figure~\ref{fig19}(c),(d)}). For the same increase in $\theta_\text{t}$, we observed $\sim$32\% drop in peak lift coefficient ({\bf Figure~\ref{fig9}(b),(d)}). In addition, average downwash velocity did not show much variation between $\theta_\text{t}$=22.5$^\circ$ and 67.5$^\circ$ ({\bf Figure~\ref{fig20}(c),(d)}). With changing $\delta$ for $\theta_\text{t}$=22.5$^\circ$, early stages of translation showed noticeable variation in $C_\text{L}$. However, during constant velocity translation, we observed little to no variation in $C_\text{L}$ for $\delta$$>$30\%. A similar trend was observed for net circulation during linear translation with increasing $\delta$, where $\Gamma_\text{LEV}$$>$$\Gamma_\text{TEV}$ in time and circulation was essentially unchanged during most of constant velocity translation across all $\delta$ ({\bf Figure~\ref{fig19}B(c),(d)}). This implies that initial wing motion helps in development of the LEV and TEV around the wing, and increasing $\delta$ decreases the strength of both the LEV and TEV. The results further imply that constant velocity translation helps in keeping both the LEV and TEV attached to the wing. However, increased downwash velocity during constant velocity translation for all $\theta_\text{r}$ and $\delta$ decreases the lift coefficient.

Increasing the overlap ($\zeta$) from $\zeta$=25\% to 100\% for $\delta$=10\% increased both $\Gamma_\text{LEV}$ and $\Gamma_\text{TEV}$, with peak net circulation being increased by $\sim$15$\%$ ({\bf Figure~\ref{fig19}(e),(f)}). Peak $C_\text{L}$ also increased by 31\% with increasing $\zeta$=25\% to 100\% ({\bf Figure~\ref{fig14}(b),(d)}), while $\overline{C_\text{L}}$ increased by 26\%  ({\bf Figure~\ref{fig14}(f)}). This substantial increase in lift coefficients is attributed to the generation of stronger LEVs for $\zeta$=100\%. This suggests that rotational acceleration during overlapping motion helps in early development of vortices. Additional acceleration from translation allowed vorticity to diffuse at both LE and TE rather than increasing its magnitude. For $\zeta$=100\%, right after 25\% of cycle time, we see a drop in $C_\text{L}$ that can be attributed to increased downwash velocity at the same instant ({\bf Figure~\ref{fig20}(f)}).

\begin{figure*}[b]
    \centering
    \includegraphics[width=0.8\textwidth]{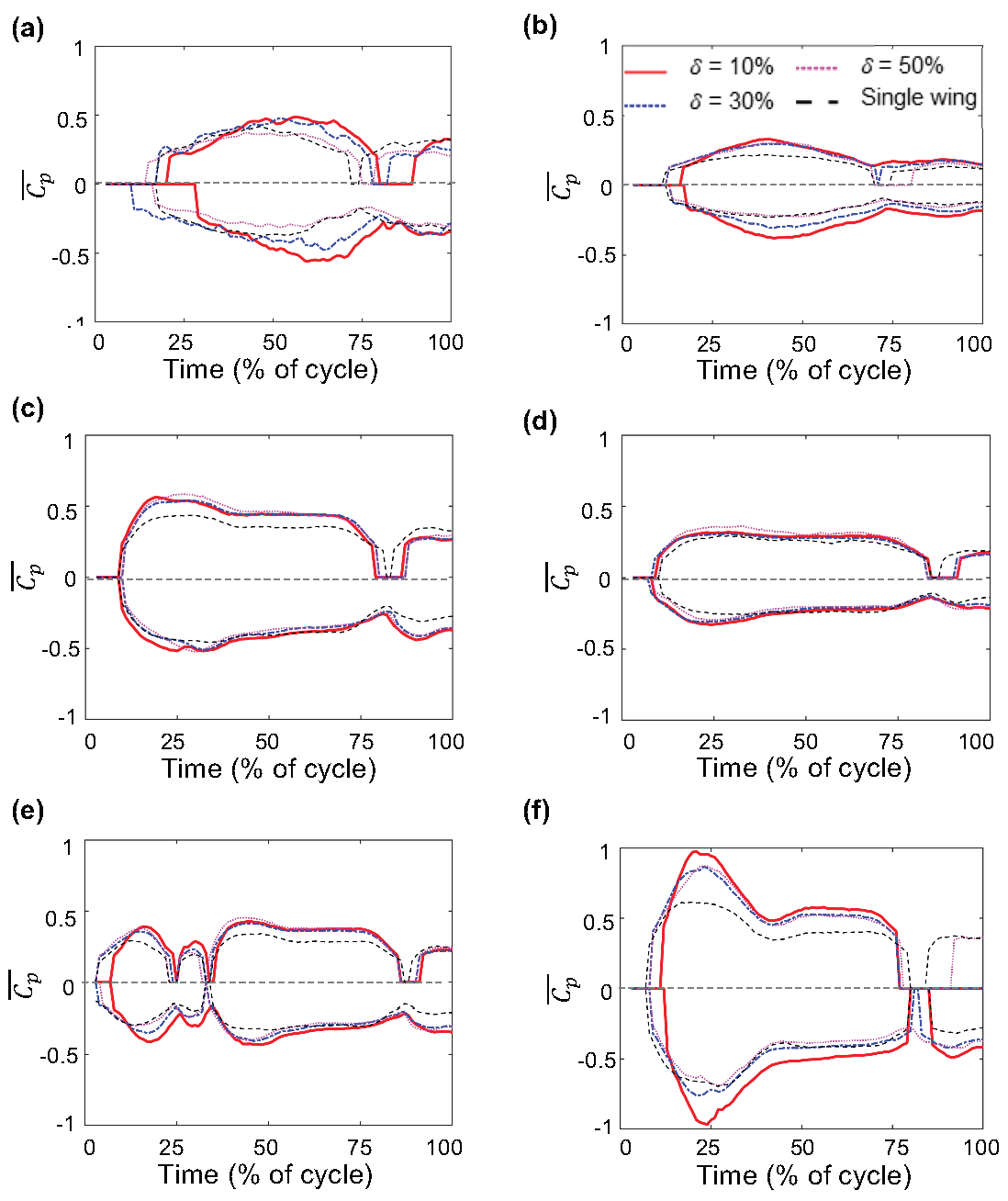}
    \caption{Time-variation of average pressure coefficient ($\overline{\textit{C}_p}$) characterizing the total dimensionless pressure distribution in the flow field, as a function of~$\delta$ and wing kinematics.~(a) and (b) show~$\overline{\textit{C}_p}$ during rotation at $\theta_\textnormal{r}$=22.5$^\circ$ and $\theta_\textnormal{r}$=67.5$^\circ$, respectively.~(c) and (d) show~$\overline{\textit{C}_p}$ during linear translation at $\theta_\textnormal{t}$=22.5$^\circ$ and $\theta_\textnormal{t}$=67.5$^\circ$, respectively.~(e) and (f) show~$\overline{\textit{C}_p}$ during combined rotation and linear translation at $\zeta$=25\% and $\zeta$=100\%, respectively. Both single bristled wing and bristled wing pairs are included. See subsection~\ref{sec:Calculated quantities} for more details on definition and calculation of $\overline{\textit{C}_p}$.
}
\label{fig21}
\end{figure*}

\subsection{Implications of pressure distribution on drag force generation}
Examining the pressure distribution on a single wing in rotation ({\bf Figure~\ref{fig7}}), we can observe that the formation of a LEV creates a low pressure region on the upper surface of the wing and a positive pressure region on the lower surface. This pressure distribution over a single rotating wing was in agreement with those reported by Cheng and Sun~\citep{XinCheng17}. For a bristled wing pair in rotation with varying $\delta$ ({\bf Figure~\ref{fig8}}), we see a negative pressure region at the top closer to the LEs and positive pressure distribution at the bottom near the TEs. In the cavity between the wings, pressure was zero to start with and becomes positive instead of negative for all $\delta$ during rotation. These results are in contrast with those of Cheng and Sun~\citep{XinCheng17}, where a negative pressure distribution was observed in between the wings at the start of fling. We suspect the positive pressure distribution in the cavity was due to strong viscous forces acting between the plates, which in turn tremendously increase drag. 

With increasing time, the positive pressure region diminished with increasing distance between the wings. The inter-wing distance in time decreases with increasing $\delta$. This suggests that smaller $\delta$ plays a crucial role in establishing the time-varying pressure field between the wings. The observed time-variation of average positive pressure coefficient ($\overline{C_\text{p}}$) was likely influenced by the positive pressure region in the cavity ({\bf Figure~\ref{fig21}(a)}). Increasing $\theta_\text{r}$ from 22.5$^\circ$ to 67.5$^\circ$ decreased the magnitude of positive pressure inside the cavity which explains the drop in $C_\text{D}$ ({\bf Figure~\ref{fig21}(a),(b)}). This drop in $C_\text{D}$ could be one of the reason for thrips to flap their wings at large rotational angles or low pitch angles ~\citep{XinCheng17} (about 20$^\circ$, equivalent to $\theta_\text{r}\approx70^\circ$). Note that pitch angle was defined relative to the horizontal in Cheng and Sun~\citep{XinCheng17}, unlike how $\theta_\text{r}$ was defined (relative to the vertical) in this study.

For smaller $\theta_\text{t}$ and for all $\delta$ that was examined here, we observed positive pressure in the cavity between the wings during early stages of linear translation of a bristled wing pair. With time, this positive pressure distribution slowly diminished as the LEs moved apart by $\sim$1.5 chord lengths. A negative pressure distribution was found to develop at the top of the wings. Interestingly, we did not see positive pressure distribution in the cavity for $\theta_\text{t}$=67.5$^\circ$ even at smaller $\delta$. We suspect that this could be due to a drop in the viscous forces acting in the cavity. Increasing $\theta_\text{t}$ was observed to decrease the magnitude of both positive and negative $\overline{C_\text{p}}$ ({\bf Figure~\ref{fig21}(c),(d)}), which explains the substantial drop in $C_\text{D}$ for larger $\theta_\text{t}$. From a recent study examining thrips wing kinematics~\citep{YuZhu19}, it was found that they operate at large $\theta_\text{t}$ values, i.e., they pitch their wings to very low angles (about 30$^\circ$, equivalent to $\theta_\text{t}$= 60$^\circ$) at the start of translation.

Similar to rotation and linear translation, we observed the formation of positive pressure region in the cavity between the wings during initial stages of wing motion for all $\zeta$ and all $\delta$ values. This positive pressure was found to diminish once the wings started moving apart. The distance between the wings where positive pressure started to diminish was found to depend on wing velocity and $\delta$. Increasing $\zeta$ increased both positive and negative $\overline{C_\text{p}}$ ({\bf Figure~\ref{fig21}(e),(f)}), which was also observed in the force coefficients. 

{\subsection{Conclusions}
Aerodynamic forces and flow structures generated by a single bristled wing and a bristled wing pair undergoing rotation about the TE(s), linear translation at a fixed angle and their combination were investigated for varying initial inter-wing spacing at \rey=10. Irrespective of $\theta_\textnormal{r}$, $\theta_\textnormal{t}$ and $\zeta$, increasing $\delta$ in a bristled wing pair decreased drag by a larger extent as compared to lift reduction due to weakening wing-wing interaction, resulting in the wing pair behaving as two single wings. A strong positive pressure region was observed in between the wings at smaller $\delta$, necessitating a large drag force to move the wings apart. The positive pressure region diminished with increasing $\theta_\textnormal{r}$, which in turn reduced drag forces. This finding suggests that a likely reason for tiny insects to employ large rotational angles (relative to vertical) in fling~\citep{XinCheng17} is to reduce drag. Finally, we find that rotational acceleration of a bristled wing aids in early development of LEV and TEV, while constant velocity translation enables the LEV and TEV to remain attached to a wing for larger $\delta$ and promotes viscous difficusion of vorticity.

Although we examined aerodynamic performance of a bristled wing model for varying kinematics, our study is limited to 2D motion. This simplification was justified by considering the phase of flapping motion where wing-wing interaction at smaller $\delta$ is observed.~An important question that remains to be investigated is whether the trends that we observed using 2D kinematics are retained when examining 3D flapping kinematics at low \rey. A previous study by Santhanakrishnan \etal\cite{Santhanakrishnan18} reported that in the \rey~range relevant to the flight of the smallest insects (\rey$\leq$32), spanwise flow decreased and viscous diffusion increased for a revolving non-bristled elliptical wing (3D motion). It is unknown how their results would be affected by the inclusion of wing bristles and when considering realistic (3D) flapping kinematics of tiny insects. These questions will be the subject of our future studies.\\

\begin{acknowledgments}
\noindent This work was supported by the National Science Foundation (CBET 1512071 to A.S.).
\end{acknowledgments}

\section*{Data availability statement}
\noindent The data that supports the findings of this study are available within the article.

\appendix*
{\section{Modeling of wing kinematics}}\label{sec:Appendix}

As mentioned in subsection~\ref{sec:Kinematics}, we used the kinematics developed by Miller and Peskin~\cite{Miller05} in this study. We used a sinusoidal velocity profile for wing rotation. The peak angular velocity ($\omega_\text{max})$ was maintained constant for each angle of wing rotation ($\theta_\text{r}$, in radians) and given by the following equation:
\begin{equation}
	\omega_\text{max}= \frac{2\theta_\text{r} U_\text{max}}{\Delta \tau_\text{rot}c}
	\label{eq:omega_max}	
\end{equation}
\noindent where~$\Delta \tau_\text{rot}$ represents the dimensionless duration of rotational phase, $c$ is the wing chord length and $U_\text{max}$ (=0.157 m s$^{-1}$) is the maximum velocity during wing rotation and linear translation. We maintained the ratio of $\theta_\text{r}$ to $\Delta \tau_\text{rot}$ constant at 0.4514 to obtain a constant $\omega_\text{max}$. The cycle time ($T$) for each $\theta_\text{r}$ was calculated using the following equation ($T$ values provided in Table~\ref{table1}):
\begin{equation}
	T = \frac{\Delta\tau_\textnormal{rot}c}{U_\textnormal{max}}
    \label{eq:CycleTime}	
\end{equation}
 For example, when $\theta_\textnormal{r}$ = 45$^\circ=\pi/4~\text{rad}$, we obtain $\Delta \tau_\textnormal{rot}$ = ($\pi$/4*0.4514)=1.74. The corresponding cycle time, $T_{\theta_\textnormal{r}=45^\circ}$=1.74$\times$0.045$\times$1000)/0.157 m s$^{-1}$=498 ms. Rounding off to nearest multiple of 10, we obtain 500 ms. 

For wing translation at a fixed angle ($\theta_\text{t}$, in radians), we employed a trapezoidal velocity profile consisting of an acceleration phase, constant velocity phase and a deceleration phase. The dimensionless duration ($\Delta\tau$) of each of these phases were maintained constant at 1.3.~The cycle time ($T$) for each translation phase was calculated from equation~\ref{eq:CycleTime}, using $\Delta\tau$ in place of $\Delta\tau_\text{rot}$: $T$=1.3$\times$0.045$\times$1000/0.157=373.  Rounding off to nearest multiple of 10, we obtain 370 ms. Total cycle time ($T$) in translation, for each $\theta_\textnormal{t}$, is given by 3$\times$370=1110 ms ($T$ values provided in Table~\ref{table1}).

The cycle time ($T$) for varying levels of overlap ($\zeta$, ranging between 0\% and 100\%) between rotation (at $\theta_\textnormal{r}$=45$^\circ$) and start of translation ($\theta_\textnormal{t}$=45$^\circ$) was calculated using the following equation:
\begin{equation}
	T_{\zeta} = \left(\frac{100-\zeta}{100}\right)T_{\theta_\textnormal{r} = 45^\circ}+T_{\theta_\textnormal{t} = 45^\circ}
\label{eq:CycleTime_Overlap}	
\end{equation}
\noindent where $T_{\zeta}$ represents cycle time for a specific $\zeta$, $T_{\theta_\textnormal{r}=45^\circ}$ and $T_{\theta_\textnormal{t}=45^\circ}$  represents cycle time of wing undergoing rotation to $\theta_\textnormal{r}$ and translation at $\theta_\textnormal{t}$=45$^\circ$. For example, when $\zeta$=25$\%$, we obtain $T_\zeta$ = (100-25)/100$\times$500+1110 ms=1485 ms. Rounding off to nearest multiple of 10, we obtain 1490 ms. Similarly, $T$ was calculated for other $\zeta$ values (provided in Table~\ref{table1}).\\

\bibliography{aipsamp}

\end{document}